\documentclass[twocolumn,preprintnumbers,prb,amsmath,amssymb]{revtex4}

\usepackage{graphicx}
\usepackage{dcolumn}
\usepackage{bm}

\begin{document}


\title{Theory of de Haas-van Alphen Effect in Type-II Superconductors}

\author{Kouji Yasui}
\affiliation{Division of Physics, Hokkaido University, Sapporo 060-0810,
Japan}
\author{Takafumi Kita}
\homepage{http://phys.sci.hokudai.ac.jp/~kita/index-e.html}
\email{kita@phys.sci.hokudai.ac.jp}
\affiliation{Division of Physics, Hokkaido University, Sapporo 060-0810,
Japan}

\date{\today}

\begin{abstract}
Theory of quasiparticle spectra and the de Haas-van Alphen (dHvA) oscillation
in type-II superconductors
are developed based on the Bogoliubov-de Gennes equations for vortex-lattice states.
As the pair potential grows through the superconducting transition,
each degenerate Landau level in the normal state splits
into quasiparticle bands in the magnetic Brillouin zone.
This brings Landau-level broadening, which in turn leads to 
the extra dHvA oscillation damping in the vortex state.
We perform extensive numerical calculations for three-dimensional systems
with various gap structures.
It is thereby shown that
(i) this Landau-level broadening is directly connected with the average gap
at $H=0$ along each Fermi-surface orbit perpendicular to the field ${\bf H}$;
(ii) the extra dHvA oscillation attenuation is caused by the broadening 
around each extremal orbit.
These results imply that the dHvA
experiment can be a unique probe to detect band- and/or angle-dependent gap amplitudes.
We derive an analytic expression for the extra damping
based on the second-order perturbation with respect to the pair potential for
the Luttinger-Ward thermodynamic potential.
This formula reproduces all our numerical results excellently, and is used to estimate
band-specific gap amplitudes from available data on NbSe$_2$, Nb$_3$Sn, and YNi$_{2}$B$_{2}$C.
The obtained value for YNi$_{2}$B$_{2}$C is fairly different from the one 
through a specific-heat measurement,
indicating presence of gap anisotropy in this material.
C programs to solve the two-dimensional Bogoliubov-de Gennes equations are available
at \url{http://phys.sci.hokudai.ac.jp/~kita/index-e.html}.
\end{abstract}

\maketitle

\section{Introduction}
The de Haas-van Alphen (dHvA) experiment on normal metals 
has been a unique and powerful tool to 
probe their Fermi surfaces.\cite{LK55,Shoenberg84}
The main purpose of this paper is to establish theoretically
that it can even be used to detect detailed gap structures
of type-II superconductors.

Back in 1976, Graebner and Robbins discovered the dHvA oscillation in 2H-NbSe$_{2}$
persisting down through the superconducting upper critical field $H_{c2}$.\cite{GR76}
It was after 15 years later
when \={O}nuki {\em et al}.\ first reconfirmed it.\cite{Onuki92}
Since then, however, a considerable number of materials have been found
to display the dHvA oscillation in the vortex state.
They include:  
$A15$ superconductors V$_3$Si\cite{Mueller92,Corcoran94} 
Nb$_3$Sn,\cite{Harrison94} 
a borocarbide superconductor YNi$_2$B$_2$C,\cite{Heinecke95,Terashima95} 
heavy-fermion superconductors CeRu$_2$,\cite{Hedo95}
URu$_2$Si$_2$,\cite{Ohkuni97,Bergemann97}
UPd$_2$Al$_3$,\cite{Inada99} CeCoIn$_5$,\cite{Settai01} and an
organic superconductor $\kappa$-(BEDT-TTF)$_2$Cu(NCS)$_2$;\cite{Wel94}
see Refs.\ \onlinecite{Janssen98} and \onlinecite{Inada99-2}
for a recent review.
Basic features of the oscillation are be summarized as follows: 
(i) the dHvA frequencies remain unchanged through the transition;
(ii) the oscillation amplitude experience an extra attenuation;
(iii) the cyclotron mass does not change
except for strongly correlated heavy fermion systems.

It is somewhat surprising that the dHvA oscillation is observable even
in superconductors without a well-defined Fermi surface.
Many theories have been presented
to explain the persistent oscillation 
and the extra damping,\cite{Maki91,Miyake93,
Gorkov98,Wasserman94,Stephen92,Curnoe00,Dukan95,Maniv92,Mineev00,Bruun97,
Gvozdikov98,NMA95,Miller95}
which may be classified into three categories.

The first approach applies a Bohr-Sommerfeld semiclassical quantization
to either the Brandt-Pesch-Tewordt\cite{Brandt67} (BPT)
Green's function near $H_{c2}$,\cite{Maki91} 
the electron number $N$ at $H=0$,\cite{Miyake93}
or Dyson's equation at $H=0$,\cite{Gorkov98} 
for obtaining the oscillatory behavior of the magnetization.
As may be seen by this diversity of the applications,
however, there is no unique semiclassical quantization scheme 
for quasiparticles superconductors;
thus, the validity of the procedure is not clear. 
This category includes Maki's theory,\cite{Maki91}
which was later reproduced by Wasserman and Springford\cite{Wasserman94} 
by treating the BTP self-energy as the extra 
broadening factor in the normal-state thermodynamic potential
and then following Dingle's procedure.\cite{Dingle52}
However, the BTP self-energy itself is obtained 
within the quasiclassical approximation
without the Landau-level structure so that this approximation
may also be questionable.

The second approach relies on some approximate analytic solutions
for the Bogoliubov-de Gennes (BdG) equations or the equivalent
Gor'kov equations,
such as the coherent-potential approximation,\cite{Stephen92,Curnoe00}
the diagonal-pairing approximation,\cite{Dukan95}
or the Ginzburg-Landau (GL) expansion for the free energy with respect 
to the order parameter.\cite{Maniv92,Mineev00,Bruun97}
However, quantitative estimations of those approximations
are yet to be carried out. 
It should also be noted that the GL expansion necessarily integrates out 
the quasiparticle degrees of freedom, so that the physical origin
of the extra oscillation damping may be obscured in the GL approach.

The third approach solves the 
BdG equations numerically without approximations.\cite{NMA95,Miller95}
Norman {\em et al}.\ \cite{NMA95} thereby extracted an analytic formula 
for the dHvA oscillation damping 
through a fitting to their numerical data.\cite{NMA95}
However, it is a two-dimensional calculation for the isotropic
$s$-wave pairing where the number of Landau levels below the 
Fermi level is $N_{\rm F}\!\sim\! 10$ at $H_{c2}$.
As may be realized from the appearance of the quantized Hall effect,
two-dimensional systems in high magnetic fields may be qualitatively
different from three-dimensional systems.
Thus, the obtained formula may not be appropriate for 
describing real three-dimensional materials with $N_{\rm F}\!\gg\! 1$.
On the other hand, another calculation by Miller and Gy\"orffy 
for a two-dimensional lattice model \cite{Miller95}
corresponds to the low-field limit near $H_{c1}$\cite{comment1}
and may not be suitable to explain the experiments.
Moreover, calculations for lattice models have a flaw that
they cannot yield continuous magnetic oscillation
due to the commensurability condition 
between the underlying lattice and the vortex lattice.

Notice finally that most of the above theories consider only 
the isotropic $s$-wave pairing.
Especially, no numerical studies
have been performed yet for anisotropic pairings or 
three-dimensional systems.

With these observations, we will perform both
numerical and analytic calculations for three-dimensional
BdG equations with various gap structures.
They can be solved efficiently by the Landau-level-expansion method (LLX),
which was formulated for vortex-lattice states 
of arbitrary pairing symmetry \cite{Kita98-2}
and used successfully to compare low-energy quasiparticle spectra between
$s$- and $d$-wave pairings.\cite{Yasui99}
We will thereby clarify how the discrete Landau levels experience 
quantitative changes as the pair potential grows below $H_{c2}$.
Another purpose is to find out the connection between the gap anisotropy
and the extra dHvA amplitude attenuation.
Terashima {\em et al}.\ \cite{Terashima95} reported a dHvA experiment on
YNi$_{2}$B$_{2}$C where an oscillation is observed to persist down to a
field $\sim\! 0.2H_{c2}$.
On the other hand, a specific-heat experiment at $H\!=\!0$
shows a power-law behavior $\propto\! T^{3}$ 
at low temperatures,\cite{Movshovich94}
indicating existence of gap anisotropy in this material.
Indeed, Izawa {\em et al}.\ \cite{Izawa02} recently reported presence of 
four point nodes in the gap based on a thermal-conductivity measurement.
Miyake \cite{Miyake93} argued that 
point or line nodes along the extremal orbit may weaken the damping,
and proposed to use the dHvA effect as a probe for gap anisotropy.
We examine this possibility in full details and present
a quantitative theory on the issue.

This paper is organized as follows.
Section \ref{sec:formulation} provides a formulation to solve
the BdG equations for vortex-lattice states.
Section \ref{sec:2D} presents calculated quasiparticle spectra and 
the dHvA oscillation for two-dimensional systems
to clarify their basic features as well as the origin of the
extra dHvA oscillation damping in the vortex state.
In Sec.\ \ref{sec:3D}, it is demonstrated that the gap anisotropy at $H\!=\! 0$
can be detectable via the dHvA oscillation in the vortex state
based on both numerical and analytic calculations 
for three-dimensional systems with various gap structures.
Section \ref{sec:estimation} presents estimations of energy gap 
for NbSe$_2$, Nb$_3$Sn, and YNi$_{2}$B$_{2}$C using the analytic formula
obtained in Appendix \ref{App:dHvA-analytic}.
Section \ref{sec:summary} concludes the paper with a brief summary.
In Appendix \ref{App:Thermo}, we derive a convenient
expression for the thermodynamic potential.
Appendix B summarizes the expressions of basis functions and
overlap integrals used in the numerical calculations.
In Appendix C, we derive an analytic expression for the extra dHvA oscillation damping
in the vortex state.
A brief report of the contents was already presented in 
Ref.\ \onlinecite{YK01}.

\section{\label{sec:formulation}Formulation}

\subsection{Bogoliubov-de Gennes equations}

Throughout the paper we will rely on the mean-field BdG equations,
which obtain the quasiparticle wavefunctions 
${\bf u}_{s}$ and ${\bf v}^{*}_{s}$ 
with a positive eigenvalue $E_{s}>0$ by
\begin{eqnarray}
&&\hspace{-10mm}\int \! d{\bf r}_{2}\left[ \!\!
\begin{array}{rc}
\vspace{1mm}\,\,\, \underline{{\cal H}}({\bf r}_{1},
{\bf r}_{2}) & \,\,
\underline{\Delta} ({\bf r}_{1},{\bf r}_{2}) \\ 
\underline{\Delta}^{\!\dagger}({\bf r}_{1},{\bf r}_{2}) 
&\! -\underline{{\cal H}%
}^{*}({\bf r}_{1},{\bf r}_{2})
\end{array}\!
\right] \! \left[ \!
\begin{array}{c}
\vspace{2mm} \,\,{\bf u}_{s}({\bf r}_{2}) \\ 
-{\bf v}^{*}_{s}({\bf r}_{2})
\end{array}\!
\right] \hspace{-3mm}  \nonumber \\
&& \hspace{20mm}= E_{s}\! \left[ \!
\begin{array}{c}
\vspace{2mm} \,\,{\bf u}_{s}({\bf r}_{1}) \\ 
-{\bf v}^{*}_{s}({\bf r}_{1} )
\end{array}\!
\right] \, .
\label{BdG}
\end{eqnarray}
Here $\underline{\Delta}$ is the pair potential and $\underline{{\cal H}}$
denotes the normal-state Hamiltonian in the magnetic field; both are $2\!\times\!2$ 
matrices to describe the spin degrees of freedom.
The symbol $^{\dagger}$
denotes Hermitian conjugate in both the coordinate and spin variables
as $[\underline{\Delta}^{\!\dagger}({\bf r}_{1},{\bf r}_{2})]_{\sigma_{1}\sigma_{2}}
\!=\! \Delta^{\! *}_{\sigma_{2}\sigma_{1}}\!({\bf r}_{2},{\bf r}_{1})$ 
with $\sigma_{j}=\uparrow,\downarrow$. 
With this definition, we can see immediately that the matrix in Eq.\ (\ref{BdG})
is Hermitian.

We adopt the free-particle Hamiltonian for $\underline{{\cal H}}$:
\begin{equation}
\underline{{\cal H}}({\bf r}_{1},{\bf r}_{2})=\delta({\bf r}_{1}\!-\!{\bf r}_{2})
\left\{\frac{\left[-i\hbar\mbox{\boldmath $\nabla$}_{\! 2}
+\frac{e}{c}{\bf A}({\bf r}_{2})\right]^{2}}{2m_{\rm e}} 
-\varepsilon_{\rm F}\right\}\underline{1}\, ,
\label{calH}
\end{equation}
where $m_{\rm e}$, $- e$ $(e>0)$, $c$, and 
$\varepsilon_{\rm F}$ are the electron mass, the electron charge, 
the light velocity,
and the chemical potential, respectively.
We will not consider the spin paramagnetism throughout.
We also neglect the spatial variation of the
magnetic field as appropriate for the relevant high-$\kappa$ materials.
Then, the vector potential ${\bf A}$ can be expressed using the
symmetric gauge as 
\begin{equation}
{\bf A}({\bf r})=-\frac{1}{2}B\hat{\bf z}\!\times\!{\bf r} \, ,
\label{Asymm}
\end{equation}
where $B$ denotes the average flux density, and
we have chosen the field along $-\hat{\bf z}$ for convenience.

The pair potential in turn is given with respect to the quasiparticle wavefunctions as
\begin{equation}
\underline{\Delta}({\bf r}_{1},{\bf r}_{2}) = V({\bf r}_{1}\!-\!{\bf r}_{2})\, 
\underline{\Phi}({\bf r}_{1},{\bf r}_{2})\, ,
\label{pair}
\end{equation}
where $V$ denotes the interaction and $\underline{\Phi}$ is the order parameter defined by
\begin{eqnarray}
\underline{\Phi}({\bf r}_1,{\bf r}_2) &\equiv& \sum_s\, 
\left[ 
{\bf u}_s({\bf r}_1){\bf v}_s^{\rm T}({\bf r}_2)-{\bf v}_s({\bf r}_1){\bf u}_s^{\rm T}({\bf r}_2)
\right] \nonumber \\
&& \times \frac{1}{2}\tanh \frac{E_s}{2k_{\rm B}T}\, ,
\label{OP}
\end{eqnarray}
with $T$ the temperature and $^{\rm T}$ denoting the transpose. 

It is shown in Appendix \ref{App:Thermo} that 
the thermodynamic potential corresponding to Eqs.\ (\ref{BdG})-(\ref{pair})
is given by
\begin{eqnarray}
&&\hspace{-3mm}\Omega = -k_{\rm B}T\sum_{s} \ln (1\!+\!{\rm e}^{-E_{s}/k_{\rm B}T})
- \sum_{s} E_{s}\!\int\! |{\bf v}_{s}({\bf r})|^{2}\, d{\bf r}  
\nonumber \\ 
&& \hspace{4.5mm}-\frac{1}{2}\int\!\!\int
\!{\rm Tr}\,\underline{\Delta}^{\!\dagger}({\bf r}_{1},{\bf r}_{2})\,
\underline{\Phi}({\bf r}_{2},{\bf r}_{1})\,{\rm d}{\bf r}_{1}{\rm d}{\bf r}_{2} \, ,
\label{Omega}
\end{eqnarray}
where Tr denotes taking trace over spin variables.
This expression will be useful to obtain an analytic expression 
for the extra dHvA amplitude attenuation in the vortex state.
The magnetization is then calculated by
\begin{equation}
M = -\frac{\partial (\Omega/{\cal V}) }{\partial B} \, ,
\label{mag}
\end{equation}
where ${\cal V}$ is the volume of the system.

\subsection{Vortex lattices and basic vectors}

Solving the above equations for general non-uniform systems
is a formidable task. 
For lattice states, however, it can be reduced into a numerically tractable
problem using the symmetry that
they are periodic with a 
single flux quantum $\phi_{0}\!\equiv\! hc/2e$ per unit cell.
We hence define a pair of basic vectors
by
\begin{equation}
\left\{ \!\! \begin{array}{l}
\vspace{2mm}
{\bf a}_{1} \equiv (a_{1x},a_{1y},0)
\\
{\bf a}_{2} \equiv (0,a_{2},0) 
\end{array} \right.\hspace{-2mm}
\, ,
\hspace{5mm}
({\bf a}_{1}\!\times\!{\bf a}_{2})\cdot\hat{\bf z}
=\frac{\phi_{0}}{B}=\pi l_{B}^{2}\, ,
\label{a1a2}
\end{equation}
where $l_{\! B}\!\equiv\!\sqrt{\hbar c/eB}$ is the magnetic length.
The basic vectors of the corresponding reciprocal lattice
are then defined by
\begin{equation}
\left\{
\begin{array}{l}
\vspace{2mm}
{\bf b}_{1}\!\equiv \! 2({{\bf a}_{2}\!\times\!\hat{\bf z}})/{l_{B}^{2}}
\\
{\bf b}_{2}\!\equiv \! 2 ({\hat{\bf z}\!\times\!{\bf a}_{1}})/{l_{B}^{2}}
\end{array}
\right. .
\label{b1b2}
\end{equation}
We now introduce magnetic Bloch vectors for quasiparticle
eigenstates by \cite{Kita98-2}
\begin{equation}
{\bf k}\equiv \frac{\mu_{1}}{{\cal N}_{\rm f}}{\bf b}_{1}+
\frac{\mu_{2}}{{\cal N}_{\rm f}}{\bf b}_{2}  
\hspace{5mm} 
\left(-\frac{{\cal N}_{\rm f}}{4}<\mu_{j}\leq \frac{{\cal N}_{\rm f}}{4}\right) \, ,
\label{k-def}
\end{equation}
and those for the center-of-mass coordinate by
\begin{equation}
{\bf q}\equiv \frac{\mu_{1}}{{\cal N}_{\rm f}}
{\bf b}_{1}+\frac{\mu_{2}}{{\cal N}_{\rm f}}
{\bf b}_{2}\hspace{5mm} 
\left(-\frac{{\cal N}_{\rm f}}{2}<\mu_{j}
\leq \frac{{\cal N}_{\rm f}}{2}\right) \, ,
\label{q-def}
\end{equation}
where ${\cal N}_{\rm f}$ is an even integer with 
${\cal N}_{\rm f}^{2}$ denoting
the number of flux quanta in the system.
Notice that ${\bf q}$ covers an area four times as large as that of {\bf k}.

\subsection{Landau-level-expansion method (LLX)}

It has been shown \cite{Kita98-2} that the pair potential 
of the conventional Abrikosov lattice
can be expanded in two ways with respect to $({\bf r}_{1},{\bf r}_{2})$
and $({\bf R},{\bf r})\equiv(\frac{{\bf r}_{1}+{\bf r}_{2}}{2},{\bf r}_{1}\!-\!{\bf r}_{2})$
as
\begin{eqnarray}
&& \hspace{-3mm}\underline{\Delta}({\bf r}_{1},{\bf r}_{2})
\nonumber \\
&& \hspace{-3mm}=\sum_{{\bf k}\alpha}\!
\sum_{ N_{1} N_{2}}\!
\underline{\Delta}_{\, N_{1}N_{2}}^{({\bf k}p_{z})}\,
\psi_{N_{1}{\bf k}\alpha}({\bf r}_{1})\,
\psi_{ N_{2}{\bf q}-{\bf k}\alpha}({\bf r}_{2})\, 
\frac{{\rm e}^{ip_{z}(z_{1}-z_{2})}}{L}
\nonumber \\
&& \hspace{-3mm}= \frac{{\cal N}_{\rm f}}{\sqrt{2}}
\sum_{N_{{\rm c}}}\sum_{N_{{\rm r}}m p_{z}} \!(-1)^{N_{\rm r}}
\underline{\bar{\Delta}}^{(N_{{\rm c}}m)}_{N_{\rm r}p_{z}}
\psi^{({\rm c})}_{N_{{\rm c}}{\bf q}}({\bf R})
\, \psi_{N_{{\rm r}}m}^{({\rm r})}({\bf r})\frac{{\rm e}^{ip_{z}z}}{L}\, .
\nonumber \\
\label{Dexp1}
\end{eqnarray}
Here $\psi_{N{\bf k}\alpha}$ is a quasiparticle basis function
with $N$ denoting the Landau level, ${\bf k}$ 
defined by Eq.\ (\ref{k-def}),
and $\alpha$ (=1,2) signifying
signifying two-fold degeneracy of every orbital state.
On the other hand, $\psi^{({\rm c})}_{N_{{\rm c}}{\bf q}}$ and
$\psi_{N_{{\rm r}}m}^{({\rm r})}$ are basis functions for the center-of-mass
and relative coordinates, respectively, with $N_{\rm c}$ and $N_{\rm r}$
denoting the corresponding Landau levels, ${\bf q}$ defined by Eq.\ (\ref{q-def}),
and $m$ an eigenvalue for the relative orbital angular momentum operator $\hat{l}_{z}$.
The quantities $p_{z}$ and $L$ are, respectively,
the wavenumber and the system length along the $z$ direction 
parallel to the magnetic field;
we adopt a notation of using ${\bf p}$ as a {\em wavevector}
in zero field to distinguish it from the two-dimensional
magnetic Bloch vector ${\bf k}$
perpendicular to the field.
As noted in Ref.\ \onlinecite{Kita98}, an arbitrary single ${\bf q}$ suffices
to describe the conventional Abrikosov lattices due to the broken translational
symmetry of the vortex lattice.
Then the first expansion of Eq.\ (\ref{Dexp1}) tells us that, 
by choosing ${\bf q}={\bf 0}$, we get an complete analogy 
with the uniform system in that the pairing occurs 
between $({\bf k},p_{z})$ and $(-{\bf k},-p_{z})$.
Finally, the two expansion coefficients 
$\underline{\Delta}_{\, N_{1}N_{2}}^{({\bf k}p_{z})}$ and 
$\underline{\bar{\Delta}}^{(N_{{\rm c}}m)}_{N_{\rm r}p_{z}}$
are connected by
\begin{subequations}
\label{D-Dt}
\begin{eqnarray}
&&\underline{\Delta}_{N_{1}N_{2}}^{({\bf k}p_{z})}
= \frac{{\cal N}_{\rm f}}{2}
\!\sum_{N_{{\rm c}}N_{{\rm r}}}\!\langle N_{1} N_{2}| N_{{\rm c}}N_{{\rm r}} \rangle 
\!\sum_{m}\langle 2{\bf k}\!-\!
{\bf q}| m\!+\!N_{\rm r}\rangle
\nonumber \\
&&\hspace{30mm}
\times(-1)^{N_{\rm r}} \underline{\bar{\Delta}}^{(N_{{\rm c}}m)}_{N_{\rm r}p_{z}}
\, ,
\label{D-Dt1}
\\
&&\underline{\bar{\Delta}}^{(N_{{\rm c}}m)}_{N_{\rm r}p_{z}}
  =  \frac{2}{{\cal N}_{\rm f}}
\sum_{N_{1} N_{2}}\langle N_{{\rm c}}N_{{\rm r}} |N_{1} N_{2}\rangle \!
\sum_{\bf k} 
\langle m\!+\!N_{\rm r}| 2{\bf k}\!-\!
{\bf q}\rangle
\nonumber \\ && \hspace{30mm}\times
(-1)^{N_{\rm r}}
\underline{\Delta}_{N_{1}N_{2}}^{({\bf k}p_{z})}\, ,
\hspace{8mm}
\label{D-Dt2}
\end{eqnarray}
\end{subequations}
where $\langle N_{1} N_{2}| N_{{\rm c}}N_{{\rm r}} \rangle $ and $\langle 2{\bf k}\!-\!
{\bf q}| m\!+\!N_{\rm r}\rangle$
are elements of unitary matrices for the basis change, i.e.\ overlap integrals.
Their explicit expressions together with those for 
$\psi_{N{\bf k}\alpha}$, 
$\psi^{({\rm c})}_{N_{{\rm c}}{\bf q}}$, and
$\psi_{N_{{\rm r}}m}^{({\rm r})}$ are given in Appendix \ref{App:Basis}.

A great advantage of using Eq.\ (\ref{Dexp1}) is that it enables us to transform
Eq.\ (\ref{BdG}) into a numerically tractable problem,
as mentioned before.
Indeed, expanding the quasiparticle wavefunctions as 
\begin{subequations}
\label{uvExpand}
\begin{eqnarray}
&& {\bf u}({\bf r})  = \sum_{N{\bf k}\alpha  p_{z}}\!{\bf u}_{s}(N) \, 
\psi_{N{\bf k}\alpha}({\bf r})\, \frac{{\rm e}^{ip_{z}z}}{\sqrt{L}}\, ,
\label{uExpand}
\\
&& {\bf v}({\bf r})  = \sum_{N{\bf k}\alpha  p_{z}}\!{\bf v}_{s}(N)\,
\psi_{N{\bf q}-{\bf k}\alpha}({\bf r})\, \frac{{\rm e}^{-ip_{z}z}}{\sqrt{L}} \, ,
\label{vExpand}
\end{eqnarray}
\end{subequations}
Eq.\ (\ref{BdG}) is reduced to a separate matrix eigenvalue problem 
for each ${\bf k}\alpha p_{z}$, and the eigenstate is labelled by 
$s\!=\!\nu{\bf k}\alpha p_{z}\sigma$ with $\nu$ and $\sigma$
denoting the quasiparticle band and its spin, respectively.
Explicitly, Eq.\ (\ref{BdG}) becomes
\begin{equation}
\sum_{N_{2}}\left[ \!
\begin{array}{rc}
\vspace{1mm}\,\,\, \underline{{\cal H}}^{(p_{z})}_{N_{1}N_{2}} & \,\,
\underline{\Delta}^{({\bf k}p_{z})}_{N_{1}N_{2}} \\ 
\underline{\Delta}^{({\bf k}p_{z})\dagger}_{N_{1}N_{2}}
&\! -\underline{{\cal H}%
}^{(p_{z})}_{N_{1}N_{2}}
\end{array}\!
\right] \!\! \left[ \!
\begin{array}{c}
\vspace{2mm} \,\,{\bf u}_{s}(N_{2}) \\ 
-{\bf v}^{*}_{s}(N_{2})
\end{array}\!
\right]\! =E_{s}\! \left[ \!
\begin{array}{c}
\vspace{2mm} \,\,{\bf u}_{s}(N_{1}) \\ 
-{\bf v}^{*}_{s}(N_{1} )
\end{array}\!
\right] \, ,
\label{BdG2}
\end{equation}
where
$\underline{\Delta}^{({\bf k}p_{z})}_{N_{1}N_{2}}$ is given by Eq.\ (\ref{D-Dt1}),
and $\underline{\cal H}^{(p_{z})}_{N_{1}N{2}}$ is diagonal as
\begin{equation}
\underline{\cal H}^{(p_{z})}_{N_{1}N{2}}=\delta_{N_{1}N_{2}}
\left[(N_{1}+{\textstyle\frac{1}{2}})\hbar\omega_{B}
+\frac{\hbar^{2}p_{z}^{2}}{2m_{\rm e}}-\varepsilon_{\rm F}\right]
\underline{1}\, ,
\label{calH2}
\end{equation}
with $\omega_{B}\!\equiv\! {eB}/{Mc}$ the cyclotron frequency.

The self-consistency equation (\ref{pair}) are also simplified greatly.
Let us define
\begin{eqnarray}
&& V_{{\bf p}{\bf p}'} \equiv \int V({\bf r}) 
\,{\rm e}^{-{\rm i}({\bf p}-{\bf p}')\cdot{\bf r}} \, {\rm d}^{3}r \, 
\nonumber \\
&&=4\pi \sum_{\ell=0}^{\infty} 
\sum_{m=-\ell}^{\ell} \bar{V}_{\ell}(p,p') Y_{\ell m}(\hat{\bf p})
 Y_{\ell m}^{*}(\hat{\bf p}') \, ,
\label{pVp'}
\end{eqnarray}
where $Y_{\ell m}(\hat{\bf p})\!\equiv\!{\Theta}_{\ell m}(\theta_{\bf p})
\frac{1}{\sqrt{2\pi}}\exp({i m\varphi_{\bf p}})$ is 
the spherical harmonic.\cite{Schiff}
We also expand both $\underline{\Delta}$ and $\underline{\Phi}$ 
in terms of the center-of-mass and relative coordinates as the
last line of Eq.\ (\ref{Dexp1}). 
Then Eq.\ (\ref{pair}) is transformed into an equation 
for the expansion coefficients of each $(N_{{\rm c}},m)$ as
\begin{equation}
\underline{{\bar\Delta}}^{\!(N_{{\rm c}}m)}_{N_{{\rm r}}p_{z}}\!
= \frac{1}{2\pi l_{\! B}^{2}L}\!\sum_{\ell N_{{\rm r}}'p_{z}'}\!\!
\bar{V}_{\ell}(p,p')
{\Theta}_{\ell m}(\theta_{\bf p}){\Theta}_{\ell m}(\theta_{{\bf p}'})
\underline{{\bar\Phi}}^{(N_{{\rm c}}m)}_{N_{{\rm r}}' p_{z}'}
\label{pair2}
\end{equation}
with
$p\!= \! \sqrt{N_{\rm r}/l_{B}^{2}\!+\! p_{z}^{2}}$ and
$\theta_{\bf p}\!=\!\tan^{-1}\bigl(\frac{\sqrt{N_{{\rm r}}}/l_{B}}{p_{z}}\bigr)$.

Let us further assume that a single $\ell$ is relevant in Eq.\ (\ref{pVp'})
and take the corresponding $\bar{V}_{\ell}(p,p')$ in a separable form as
\begin{equation}
\bar{V}_{\ell}(p,p')=V_{\ell} \, W_{\ell}(\xi)W_{\ell}(\xi') \, ,
\label{Vsep}
\end{equation}
where $W_{\ell}(\xi)$ is some cut-off function with respect to
$\xi\!\equiv\! \hbar^{2}p^{2}/2m_{\rm e}\!-\!\varepsilon_{\rm F}$
satisfying $W_{\ell}(0)\!=\!1$.
Then can rewrite Eq.\ (\ref{pair2}) as
\begin{subequations}
\label{pair3}
\begin{equation}
\underline{{\bar\Delta}}^{\!(N_{{\rm c}}m)}_{N_{{\rm r}}p_{z}}=
\underline{\tilde\Delta}^{(N_{{\rm c}}m)}
W_{\ell}(\xi)
\Theta_{\ell m}(\theta_{\bf p})\, ,
\label{pair3a}
\end{equation}
with $\xi\!=\! \hbar^{2}(N_{\rm r}/l_{B}^{2}\!+\! p_{z}^{2})/2m_{\rm e}\!-\!\varepsilon_{\rm F}$
and
\begin{equation}
\underline{\tilde\Delta}^{(N_{{\rm c}}m)}
= \frac{V_{\ell}}{2\pi l_{\! B}^{2}L}\!\sum_{ N_{{\rm r}}'p_{z}'}\!\!
W_{\ell}(\xi'){\Theta}_{\ell m}(\theta_{{\bf p}'})\, 
\underline{{\bar\Phi}}^{(N_{{\rm c}}m)}_{N_{{\rm r}}' p_{z}'} \, .
\label{pair3b}
\end{equation}
\end{subequations}
Thus, we only need a self-consistent solution 
for a set of discrete parameters 
$\{\underline{\tilde\Delta}^{(N_{{\rm c}}m)}(T,B)\}$
through Eqs.\ (\ref{BdG2}) and (\ref{pair3}) 
using Eq.\ (\ref{D-Dt}).

It has been shown \cite{Ryan93,Kita98} 
that retaining a few $N_{\rm c}$'s,
e.g., $N_{\rm c}\!=0,6,12$ for the hexagonal lattice,
is sufficient to describe the vortex lattices of $H\!\agt\!0.05H_{c2}$.
Thus, the original problem of obtaining self-consistency 
for $\underline{\Delta}({\bf r}_{1},{\bf r}_{2})$
at all space points is now reduced to
the one for a few expansion coefficients 
$\{\underline{\tilde\Delta}^{(N_{{\rm c}}m)}(T,B)\}$.
This situation is analogous to the zero-field case
where a single parameter $\Delta_{0}(T)$
specifies the pair potential.

The linearized self-consistency equation
is obtained by substituting into Eq.\ (\ref{pair3b}) the expression of
$\underline{\bar\Phi}^{(N_{{\rm c}}m)}_{N_{{\rm r}}p_{z}}$ 
linear in the pair potential:\cite{Kita98-2}
\begin{eqnarray}
&&\hspace{-7mm}\underline{\bar\Phi}^{(N_{{\rm c}}m)}_{N_{{\rm r}}p_{z}}
=- \frac{1}{2}\sum_{N_{1}N_{2}} 
\langle  N_{\rm c}N_{\rm r}|N_{\rm 1}N_{\rm 2}\rangle 
\,\frac{\tanh\frac{\xi_{1}}{2T}\!+\!
\tanh\frac{\xi_{2}}{2T}}{\xi_{1}\!+\!\xi_{2}}
\nonumber \\
&& \hspace{-3mm} \times
\sum_{n}(-1)^{n}
\langle  N_{\rm 1}N_{\rm 2}|  N_{\rm c}\!+\!n \,  N_{\rm r}\!-\!n \rangle \,
\underline{\bar\Delta}^{(N_{\rm c}+n\, m+n )}_{N_{{\rm r}}\!-\! n\, p_{z}} \, .
\label{Phi-linear}
\end{eqnarray}
Equation (\ref{pair3}) with Eq.\ (\ref{Phi-linear}) determines the mean-field 
$H_{c2}(T)$, i.e.\ $T_{c}(H)$.
If we use the asymptotic expression (\ref{B_NN'asym}) for the overlap integral
$\langle  N_{\rm c}N_{\rm r}|N_{\rm 1}N_{\rm 2}\rangle$
and replace the sum over $N_{1}$ by the integral over $x\!\equiv\!(N_{1}\!-\!N_{2})/
\sqrt{2(N_{\rm c}\!+\! N_{\rm r})}$, 
we reproduce the smooth quasiclassical $H_{c2}^{\rm quasi}(T)$
obtained, for example, for the $s$-wave pairing by Helfand and Werthamer.\cite{HW66}

Finally, Eq.\ (\ref{pair}) for two-dimensional systems can be
transformed similarly.
It is also obtained from Eqs.\ (\ref{pVp'})-(\ref{pair3}) by replacing
$V_{\ell}(p,p')\!\rightarrow\! V^{(m)}(p,p')$, extending the summation over $m$
in Eq.\ (\ref{pVp'}) from $-\infty$ to $\infty$,
and finally restricting the summation over $\ell$ and $p_{z}$ only to $\ell\!=\!0$
and $p_{z}\!=\!0$, respectively.

\section{\label{sec:2D}Two-Dimensional calculations}

We first consider a couple of two-dimensional models
and perform fully self-consistent calculations.
Our purposes in this section are summarized as follows:
(i) to clarify the essential features of the
results by self-consistent calculations;
(ii) to see whether point nodes in the gap really
enhances the dHvA signals as Miyake claims.\cite{Miyake93}

\subsection{Models}

The one-particle Hamiltonian (\ref{calH}) yields an isotropic 
Fermi surface
specified by a unit vector $\hat{\bf p}\!=\!(\cos\varphi_{\bf p}, 
\sin\varphi_{\bf p})$.
As for the pairing interaction (\ref{pVp'}),
we adopt the following models:
\begin{equation}
V_{{\bf p}{\bf p}'}=
\left\{ 
\begin{array}{l}
\vspace{2mm} \! V_{0}\, W(\xi)W(\xi')  \\ 
\!V_{2}\, W(\xi)(\hat{\bf p}_{x}^{2}\!-\!
\hat{\bf p}_{y}^{2})\, W(\xi') (\hat{\bf p}_{x}^{\,\prime 2}\!-\!
\hat{\bf p}_{y}^{\,\prime 2}) 
\end{array}
\right. ,
\label{pVp'2}
\end{equation}
where
\begin{equation}
W(\xi)=\exp\!\left[-
\frac{1}{2}\!\left(\frac{\xi}{\hbar\omega_{\rm D}}\right)^{4}\,\right] 
\label{Wcut}
\end{equation}
is a smooth cut-off function with $\omega_{\rm D}$ a cut-off frequency.
The second model of Eq.\ (\ref{pVp'2}) is beyond the original isotropic interaction 
(\ref{pVp'}),
but it is convenient for the above-mentioned purposes.
In zero field, the two interactions yield the $s$- and 
$d_{x^2-y^2}$-wave gaps as
\begin{equation}
{\underline \Delta}_{{\bf p}}=
\left\{ 
\begin{array}{ll}
\vspace{2mm} \! \Delta_{0} W(\xi)\,i {\underline \sigma}_{2} & \mbox{: $s$-wave} \\ 
\!\Delta_{0} W(\xi)(\hat{\bf p}_{x}^{2}\!-\!
\hat{\bf p}_{y}^{2})\,i{\underline \sigma}_{2} & \mbox{: $d_{x^2-y^2}$-wave}
\end{array}
\right. ,
\label{gap2D}
\end{equation}
respectively. 
The corresponding $\underline{{\bar\Delta}}^{\!(N_{{\rm c}}m)}_{N_{{\rm r}}}$
of Eq.\ (\ref{pair3}) for ${\bf B}\!\parallel\!\hat{\bf z}$ is given by
\begin{subequations}
\label{pair2D}
\begin{equation}
\underline{{\bar\Delta}}^{\!(N_{{\rm c}}m)}_{N_{{\rm r}}}
= \left\{ 
\begin{array}{ll}
\vspace{2mm}  \! {\tilde\Delta}^{(N_{\rm c})} W(\xi) \, 
\delta_{m0} \,i {\underline \sigma}_{2} \\ 
\! {\tilde\Delta}^{(N_{\rm c})} W(\xi)  
 {\displaystyle\frac{\delta_{m2}\!+\!\delta_{m-2}}{2}}
\,i{\underline \sigma}_{2} 
\end{array}
\right. ,
\label{pair2D1}
\end{equation}
with $\xi\!\equiv\! \hbar^{2}N_{\rm r}/2m_{\rm e}l_{B}^{2}\!-\!\varepsilon_{\rm F}$ and
\begin{equation}
\tilde\Delta^{(N_{{\rm c}})}
= \left\{
\begin{array}{l}
\vspace{0mm}
\displaystyle
\!\frac{V_{0}}{4\pi l_{\! B}^{2}}\sum_{ N_{{\rm r}}'}\!
W_{\ell}(\xi')\,
{\bar\Phi}^{(N_{{\rm c}},0)}_{N_{{\rm r}}'} 
\\
\displaystyle
\!\frac{V_{2}}{4\pi l_{\! B}^{2}}\sum_{ N_{{\rm r}}'}\!
W_{\ell}(\xi')\,\,\frac{
{\bar\Phi}^{(N_{{\rm c}},2)}_{N_{{\rm r}}'} \!+\! 
{\bar\Phi}^{(N_{{\rm c}},-2)}_{N_{{\rm r}}'} }{2}
\end{array}
\right.
\, .
\label{pair2D2}
\end{equation}
\end{subequations}
Here we have adopted a normalization for ${\tilde\Delta}^{(N_{\rm c})}$
different from Eq.\ (\ref{pair3}) so that ${\tilde\Delta}^{(N_{\rm c})}$
acquires a direct correspondence to the maximum
gap $\Delta_{0}$ in Eq.\ (\ref{gap2D});
the factor $\frac{1}{2}$ in the second case stems from 
$\cos2\varphi_{\bf p}$ in Eq.\ (\ref{gap2D}).

We have chosen $V_{\ell}$ in Eq.\ (\ref{pVp'2}) as
\begin{equation}
g_\ell\!\equiv\!-N(0)V_{\ell}\!=\!0.5 \, ,
\label{g_l}
\end{equation}
where $N(0)\!=\!m_{e}/2\pi\hbar^{2}$ is the density of states per spin at the Fermi level.
Another important parameter is the zero-temperature coherence length defined by
\begin{eqnarray}
\xi_{0}\equiv {\hbar v_{\rm F}}/{\Delta_{0}} \, ,
\label{kFxi0}
\end{eqnarray}
with $v_{\rm F}$ the Fermi velocity.
We have adopted $p_{\rm F}\xi_{0}\!=\! 5$ for our calculations.
The above two quantities fix our models completely;
the cut-off $\hbar \omega_{\rm D}$ in Eq.\ (\ref{Wcut}) 
and $T_{c}(B\!=\!0)$ are calculated using the gap equation.

It should be noted that choosing $p_{\rm F}\xi_{0}$ also determines
the following quantities:
(i) the ratio $\hbar\omega_{H_{c2}^{\rm quasi}}/k_{\rm B}T_{c}$, where
$\hbar\omega_{H_{c2}^{\rm quasi}}$ is
the zero-temperature cyclotron energy at the quasiclassical
upper critical field $H_{c2}^{\rm quasi}$;
(ii) the number $N_{\rm F}$ of the Landau levels 
below the Fermi level at $H_{c2}^{\rm quasi}$.
Indeed, using the usual cut-off model 
$W(\xi)\!=\!\theta(\hbar\omega_{\rm D}-|\xi_{\bf p}|)$
with $\theta$ the step function, 
we obtain the following results for the $s$-wave pairing:
\begin{eqnarray}
\left\{
\begin{array}{l}
\vspace{2mm}
\hbar\omega_{H_{c2}^{\rm quasi}}/k_{\rm B}T_{c} =  6.28/p_{\rm F}\xi_{0} \, ,\\
N_{\rm F} \equiv \varepsilon_{\rm F}/\hbar\omega_{H_{c2}^{\rm quasi}}=
0.140 (p_{\rm F}\xi_{0})^{2} \, .
\end{array}
\right.
\label{Free-Particle}
\end{eqnarray}
To reproduce the experimental situation $\hbar\omega_{H_{c2}^{\rm quasi}}/k_{\rm B}T_{c}\!=\!
1\!\sim\!3$ within the present model,
we should go into the quantum limit 
$p_{\rm F}\xi_{0}\!=\! 6\!\sim\!2$, but we then have $N_{\rm F}\!=\! 5\!\sim\! 1$.
In real materials, however, $\hbar\omega_{H_{c2}^{\rm quasi}}/k_{\rm B}T_{c}$
and $p_{\rm F}\xi_{0}$ are apparently independent parameters
due to effects not covered by the free-particle model 
such as the energy band structure.
Indeed,  
$p_{\rm F}\xi_{0}$ is of the order of $30$ (NbSe$_{2}$) or even larger,
whereas $\hbar\omega_{H_{c2}^{\rm quasi}}/k_{\rm B}T_{c}\!=\! 1\!\sim\!3$;
see Table \ref{table1} below.
Also, $N_{\rm F}\!\gg\! 1$ for those materials.
The failures to describe these situations are
among the main difficulties of the free-particle model of Eq.\ (\ref{calH}).

Motivated by these observations, we also perform another 
calculations with much more Landau levels below the Fermi level.
This is achieved by including the effect of
the band dispersion.
Specifically, we apply the Onsager-Lifshiz (OL) quantization scheme
to $\underline{{\cal H}}$ of Eq.\ (\ref{BdG}),
i.e.\ the procedure which has been very successful 
for describing the dHvA oscillations in the normal state.\cite{LK55,Shoenberg84}
Given the density of states per spin $N(\varepsilon)$
and the average flux density $B$,
the $N$th Landau level $\varepsilon_{N}$ ($N\!=\! 0,1,2,\cdots$) is determined by
\begin{eqnarray}
2\left(\! N\!+\!\frac{1}{2} \right)\! \frac{\hbar^{2}}{l_{B}^{2}} = 
4\pi\hbar^{2}\!\!\int_{0}^{\varepsilon_{N}}\!\!
N(\varepsilon')\,
d\varepsilon' \, .
\label{OL}
\end{eqnarray}
We fix $\underline{\cal H}_{N_{1}N_{2}}^{(p_{z})}$ of Eq.\ (\ref{BdG2})
in this way assuming it is diagonal.
In contrast, we use the same expression 
for $\underline{\Delta}_{N_{1}N_{2}}^{({\bf k}p_{z})}$ of Eq.\ (\ref{BdG2})
as the free-particle case.
Finally, we choose $\xi\!=\! \varepsilon_{N_{\rm r}/2}-\varepsilon_{\rm F}$ 
in Eq.\ (\ref{pair2D}) based on the consideration 
of the free-particle model.\cite{Kita98-2}
With these prescriptions together with the transformation (\ref{D-Dt}), 
the coupled equations (\ref{BdG2}) and (\ref{pair2D}) are defined unambiguously.
We adopt the model density of states:
\begin{eqnarray}
N(\varepsilon)= \frac{m_{\rm e}}{2\pi\hbar^{2}}
\left(1+\frac{\alpha\Gamma}{\varepsilon^{2}+\Gamma^{2}}\right) ,
\label{DOS}
\end{eqnarray}
and choose the numerical constants $(\alpha,\Gamma)\!=\!(2.1,2.7)$ and $(5.0,1.0)$
for the $s$- and $d$-wave models of Eq.\ (\ref{gap2D}), respectively.
We also use Eq.\ (\ref{g_l}) for the pairing interaction
and fix $\hbar\omega_{\rm D}\!=\!0.5\varepsilon_{\rm F}$ in Eq.\ (\ref{Wcut}).
We thereby obtain
$\hbar\omega_{H_{c2}^{\rm quasi}}\!\approx\!k_{\rm B}T_c$ at $T\!=\!0$
and $N_{\rm F}\!\sim\!30$ at $H\!=\!H_{c2}^{\rm quasi}$,
which describe the experimental situation much
better than the free-particle model.

\subsection{Numerical procedures}
\label{NumProc2D}

Coupled equations (\ref{BdG2}) and (\ref{pair2D}) are solved iteratively
with the help of the transformation (\ref{D-Dt})
to obtain self-consistent $\{{\tilde\Delta}^{(N_{\rm c})} \}$'s
and the corresponding quasiparticle eigenstates. 
The hexagonal (square) vortex lattice is assumed 
for the $s$-wave ($d_{x^2-y^2}$-wave) model, 
as expected theoretically
in high magnetic fields\cite{Won96,Ichioka99}
and observed recently 
in La$_{1.83}$Sr$_{0.17}$CuO$_{4+\delta}$.\cite{Gilardi02}
It should be noted, however, that the precise lattice structure is
not important for the theory of the dHvA oscillation in superconductors.
We set ${\bf q}\!=\!\frac{1}{2}({\bf b}_{1}\!+\!{\bf b}_{2})$ in the relevant equations
so that a core of the pair potential is located at the origin
${\bf R}\!=\!{\bf 0}$.\cite{Kita98}
An advantage of this choice is that 
the corresponding quasiparticle energies
have the rotational symmetry 
of the hexagonal (square) lattice around ${\bf k}\!=\!{\bf 0}$;
other choices would shift the rotation axis from the origin.
We then perform calculations of Eqs.\ (\ref{BdG2}) and (\ref{pair2D})
for a set of discrete ${\bf k}$'s defined by Eq.\ (\ref{k-def}),
where ${\cal N}_{\rm f}$ is chosen as a multiple of $12$
to include all the high-symmetry points 
$\Gamma$, $M$, and $K$ ($\Gamma$, $X$, and $M$) 
of the hexagonal (square) lattice.
Three different values ${\cal N}_{\rm f}\!=\! 12,24,36$
are used to see the size dependence, and
it has been checked that the results do not differ for the three cases. 
The hexagonal (square) symmetry
enables us to restrict the summation over ${\bf k}$
into approximately $1/12$ ($1/8$) area of the Brillouin zone.
Thus, the calculations can be reduced greatly
with due care on the degeneracy of high-symmetry points.
Finally, the obtained eigenvalues and eigenstates are substituted
into Eq.\ (\ref{Omega}) 
to calculate the magnetization by Eq.\ (\ref{mag}).
All the calculations are performed at $T\!=\!0.1T_{c}$.

\begin{figure}[t]
\includegraphics[width=0.45\linewidth]{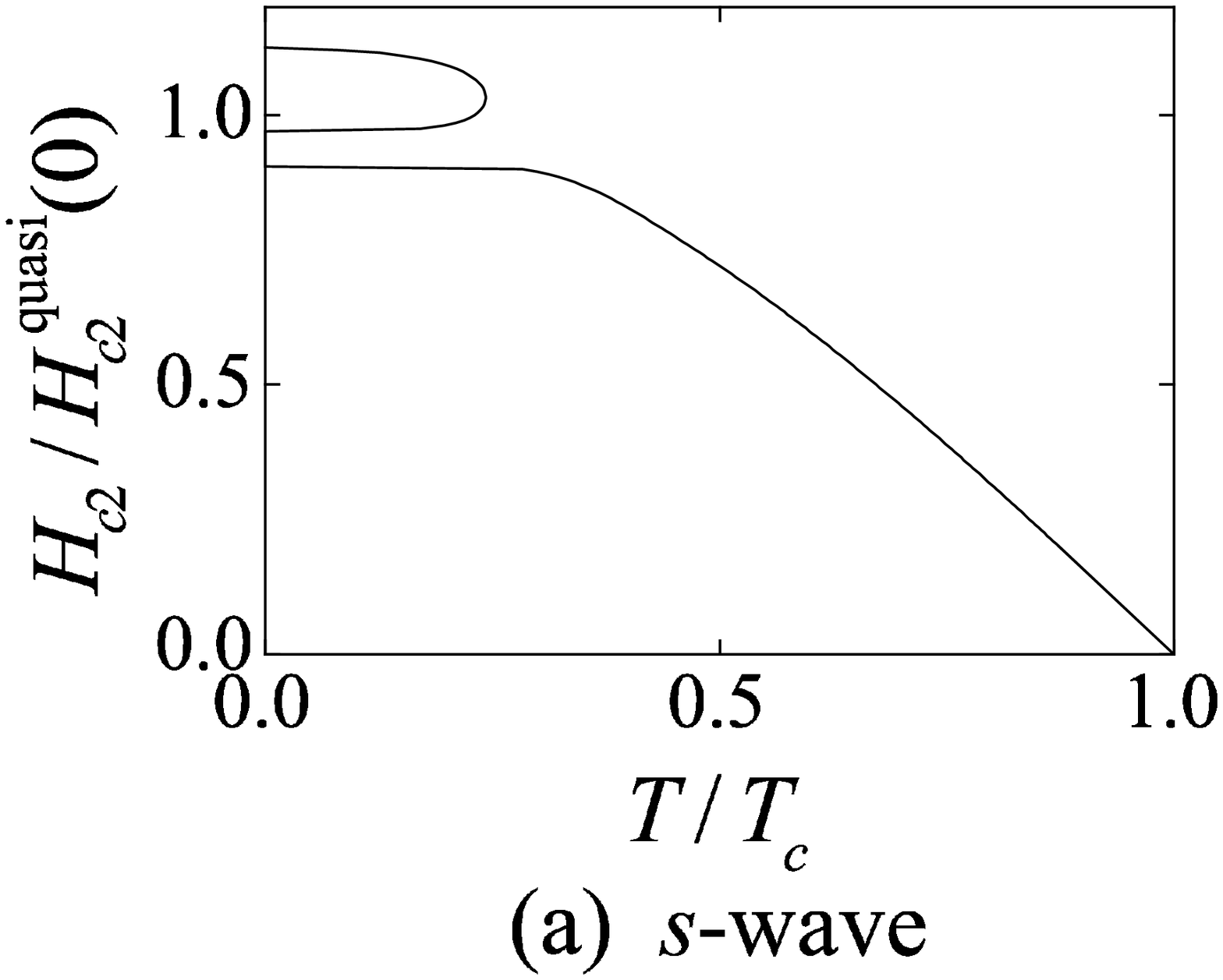}
\hspace{2mm}
\includegraphics[width=0.45\linewidth]{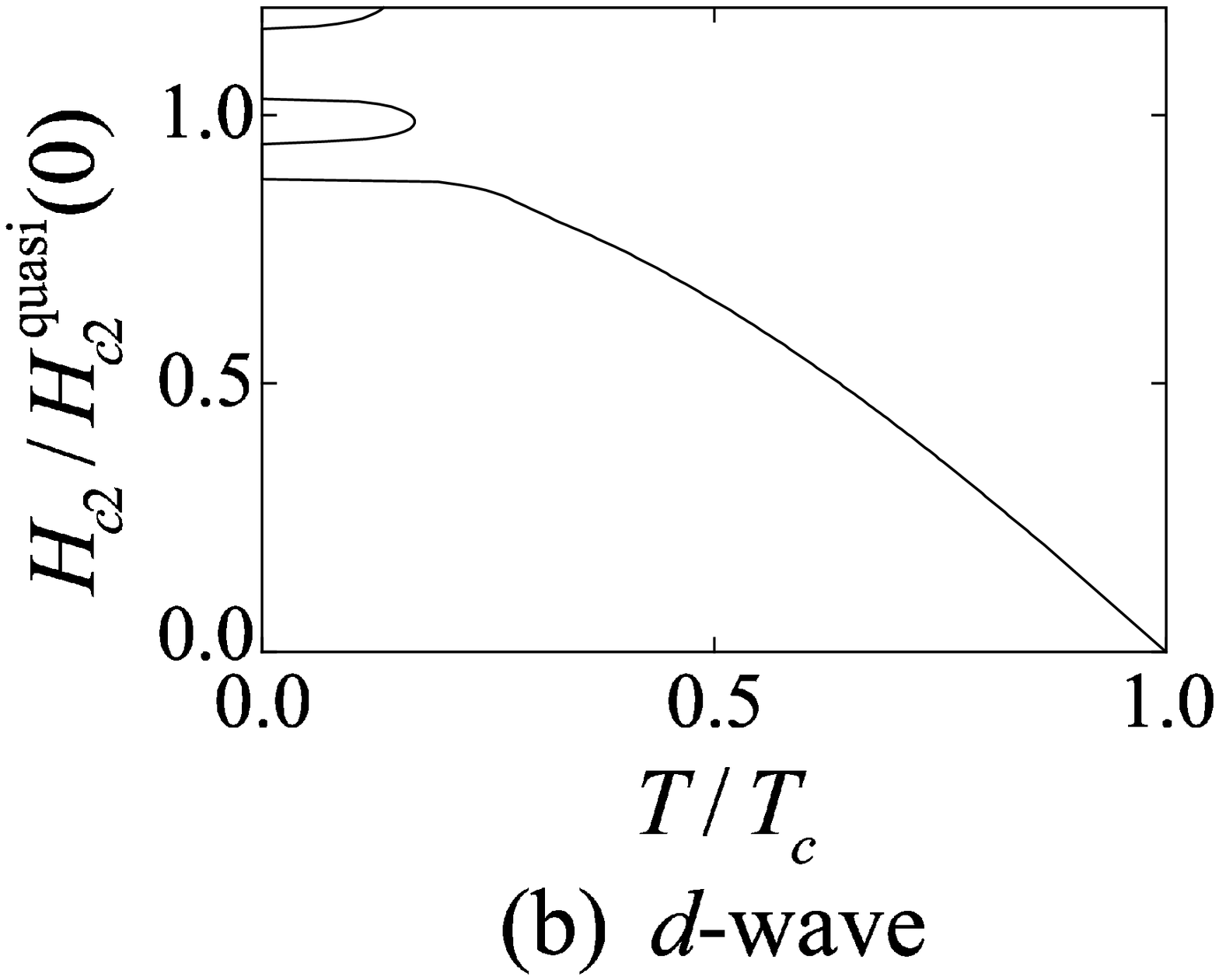}
\caption{The upper critical field $H_{c2}$ as a function of 
$T$ for (a) $s$-wave and (b) $d$-wave
of Eq.\ (\ref{gap2D}).
Here $p_{\rm F}\xi_{0}\!=\! 5$, and $H_{c2}$ is normalized by 
the quasiclassical upper
critical field $H_{c2}^{\rm quasi}(T\! =\! 0)$. }
\label{fig:Hc2-self}
\end{figure}
\begin{figure}[h]
\includegraphics[width=0.99\linewidth]{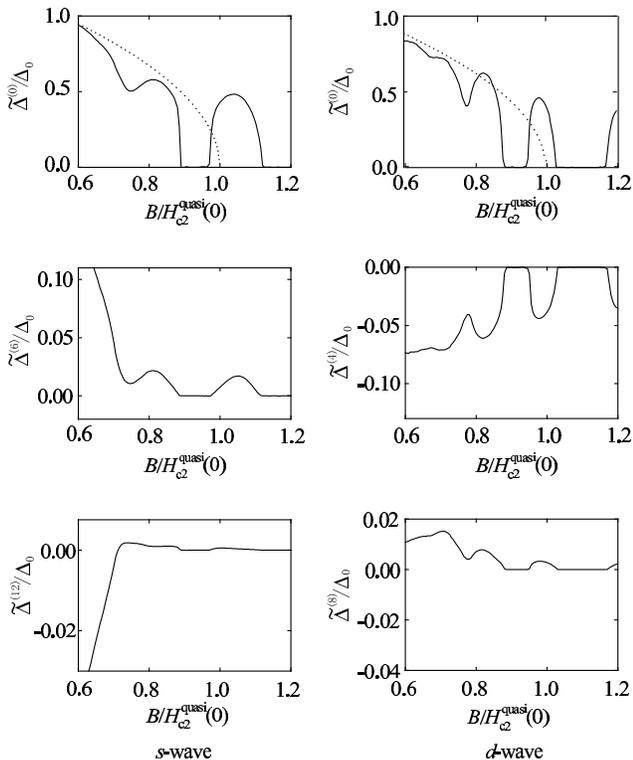}
\caption{The expansion coefficients $\tilde{\Delta}^{(N_{\rm c})}$
in Eq.\ (\ref{pair2D}) 
as a function of $B$ for the $s$-wave (first column) and the $d$-wave (second column)
with $T\!=\!0.1T_{c}$ and $p_{\rm F}\xi_{0}\!=\! 5$.
The dotted lines in the first row signify the square-root
behavior expected from the quasiclassical theory. }
\label{fig:delta-self}
\end{figure}
\begin{figure}[t]
\includegraphics[width=0.95\linewidth]{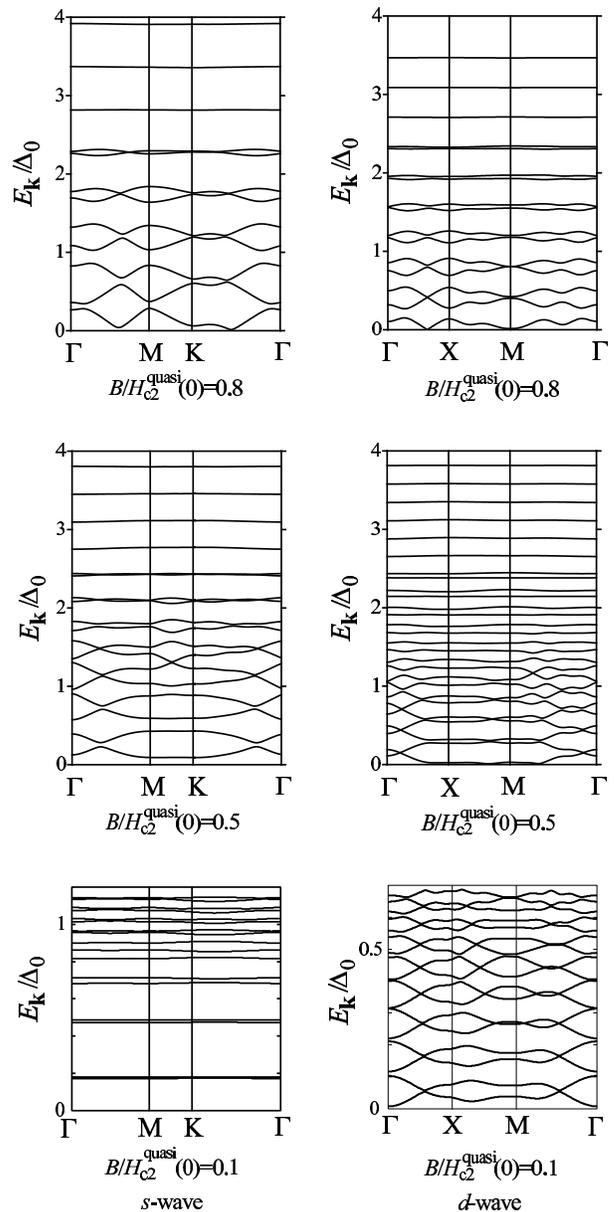}
\caption{Quasiparticle dispersion in the magnetic Brillouin zone
for the $s$-wave hexagonal lattice (first column) and the $d$-wave square lattice
(second column). Here $p_{\rm F}\xi_{0}\!=\! 5$,  $T\!=\!0.1T_{c}$, 
and $B/H_{c2}^{\rm quasi}(0)$ is equal to 
$0.8$, $0.5$, and $0.1$ from top to bottom, respectively}
\label{fig:disp-self}
\end{figure}

\subsection{Results}

The above self-consistent procedure is known to give rise to oscillatory singular 
behaviors in both $H_{c2}$ and
${\tilde\Delta}^{(N_{{\rm c}})}$ in the field range
where the dHvA oscillation persists.\cite{Rasolt87,MacDonald92,Rasolt92}
Figure \ref{fig:Hc2-self} displays $H_{c2}(T)$ 
calculated self-consistently for the $s$- and $d$-wave models;
it is normalized by the quasiclassical upper critical field 
$H_{c2}^{\rm quasi}(T\!=\!0)$.
An oscillatory behavior sets in around $T\!\alt\! 0.2T_{c}$,
and $H_{c2}$ deviates substantially from the smooth Helfand-Werthamer behavior\cite{HW66}
predicted by the quasiclassical theory.
The number of the Landau levels below the Fermi level
is $N_{\rm F}\!\sim \! 10$ around $H_{c2}^{\rm quasi}(0)$,
which is considerably smaller than those for the real materials.
Figure \ref{fig:delta-self} shows ${\tilde\Delta}^{(N_{\rm c})}$
as a function of $B$
at $T\!=\!0.1T_{c}$ for the $s$-wave hexagonal lattice (first column) 
and the $d$-wave square lattice (second column);
they are real and finite only for $N_{\rm c}\!=\!0,6,12,\cdots$ $(0,4,8,\cdots)$ 
for the hexagonal (square) lattice,\cite{Ryan93,Kita98} as already mentioned.
We observe that ${\tilde\Delta}^{(N_{\rm c})}$'s
are also singular, and the dominant ${\tilde\Delta}^{(0)}$
component cannot be described by the square-root behavior
near $H_{c2}^{\rm quasi}(0)$ expected from the quasiclassical theory.
However, those singular behaviors disappear gradually as $B$ decreases.

\begin{figure}[t]
\includegraphics[width=0.99\linewidth]{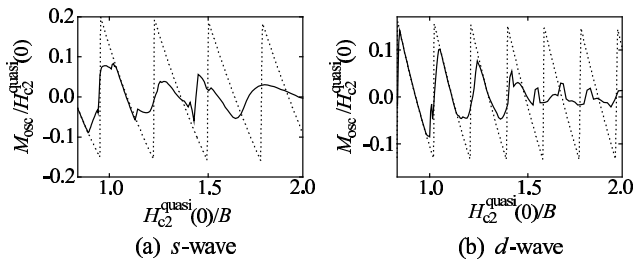}
\caption{Oscillatory part of magnetization $M_{\rm osc}$ for (a) the $s$-wave
and (b) the $d$-wave, 
over $0.8\!\leq\!H_{c2}^{\rm quasi}(0)/B\!\leq \! 2.0$ 
($1.2 \!\geq\! B/H_{c2}^{\rm quasi}(0)\!\geq \! 0.5$)
with $T\!=\!0.1T_{c}$ and  $p_{\rm F}\xi_{0}\!=\! 5$. 
The dotted lines are the 
curves of the corresponding normal state.}
\label{fig:Mosc-self}
\end{figure}

Figure \ref{fig:disp-self} displays the quasiparticle energies in the magnetic Brillouin zone
for the $s$-wave hexagonal lattice (first column) 
and the $d$-wave square lattice (second column)
at $B/H_{c2}^{\rm quasi}(0)\!=\!0.8$, $0.5$, and $0.1$.
At $B/H_{c2}^{\rm quasi}(0)\!=\!0.8$, we already observe large dispersion 
for $E\!\alt\!2\Delta_{0}$ where the pair potential is effective.
In contrast, the flat Landau-level structure remains for $E\!\agt\!2\Delta_{0}$
where the pair potential vanishes in the present cut-off model of Eq.\ (\ref{Wcut}).
Thus, the dispersion is caused clearly by the scattering from the growing pair potential, 
and as will be discussed below, it is the origin of the
extra dHvA oscillation damping in the vortex state.
We also notice that, for $B/H_{c2}^{\rm quasi}(0)\!\agt \! 0.5$,
almost no qualitative difference can be seen between the $s$- and $d$-wave cases.
At a lower field of $B/H_{c2}^{\rm quasi}(0)\!=\! 0.1$, however,
a marked difference grows around $E\!\alt \!\Delta_{0}$.
The $s$-wave energy bands of $E\!\alt \! 0.7\Delta_0$ 
are flat and occur in pairs with the level
spacing of the order of $\Delta_0^2/\varepsilon_{{\rm F}}$. 
As already pointed out by Norman {\em et al}.,\cite{NMA95}
these corresponds to the bound core states of an isolated vortex
with little tunneling probability between adjacent cores. 
In contrast, the $d$-wave bands in the same region
are densely packed with large dispersion, indicating the extended nature of
the corresponding quasiparticle wavefunctions.
From this comparison, we conclude that
no bound states exist for the $d$-wave model even in the zero-field limit of
an isolated vortex, in agreement with the result of Franz and Te\v sanovi\'c.\cite{Franz98}
This difference in the low-energy dispersion at low fields 
was already reported in Ref.\ \onlinecite{Yasui99}.

\begin{figure}[t]
\includegraphics[width=0.99\linewidth]{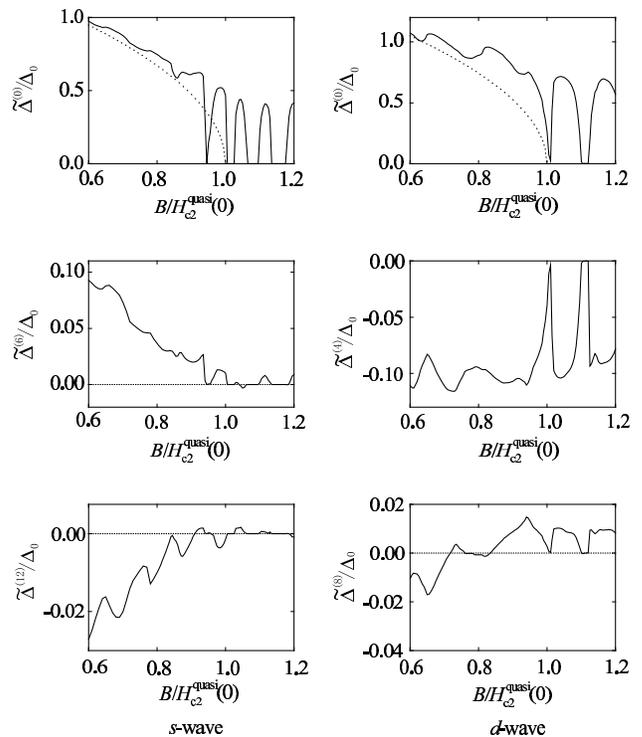}
\caption{The expansion coefficients $\tilde{\Delta}^{(N_{\rm c})}$
in Eq.\ (\ref{pair2D})
as a function of $B$ for the $s$-wave (first column) and the $d$-wave (second column).
Here  $T\!=\!0.1T_{c}$, and the non-quadratic dispersion given by Eq.\ (\ref{DOS}) is used.
The dotted lines in the first row signify the square-root
behavior expected from the quasiclassical theory. }
\label{fig:delta-self-nonquadra}
\end{figure}

Figure \ref{fig:Mosc-self} shows oscillatory part
of the magnetization $M_{\rm osc}$ calculated numerically by Eq.\ (\ref{mag}),
where curves of the corresponding normal state
are also plotted for comparison. 
The damping starts from above $H_{c2}^{\rm quasi}(0)$
where ${\tilde\Delta}^{(0)}$ is already finite as in Fig.\ \ref{fig:delta-self},
and develops rapidly as ${\tilde\Delta}^{(0)}$ grows in decreasing $B$.
Thus, the mean-field theory predicts that the dHvA oscillation comes together
with the oscillatory singular behaviors in $H_{c2}$ and 
${\tilde\Delta}^{(N_{\rm c})}$. 
Combined with the energy dispersion given in Fig.\ \ref{fig:disp-self},
we are now able to attribute the origin of the extra damping unambiguously to
the Landau-level broadening due to the pair potential.
The oscillations are rather irregular in both the $s$-wave and $d$-wave
cases, in accordance with the singular behaviors of $\Delta^{(N_{\rm c})}$ 
in Fig.\ \ref{fig:delta-self}.
We also see no qualitative difference between the two cases.

However, 
the free-particle model has several inappropriate points
as discussed already around Eq.\ (\ref{Free-Particle}).
For example, the number of Landau levels below the Fermi level $N_{\rm F}$ 
is necessarily $N_{\rm F}\!\sim\! 10$ at $H_{c2}^{\rm quasi}$
for $p_{\rm F}\xi_{0}\!=\! 5$,
which is much smaller than the values of the materials
displaying the dHvA oscillation.
Hence the above numerical results may not be sufficient
to say anything quantitative about the dHvA
attenuation or the differences between the $s$- and $d$-wave cases.
We have thus performed another calculations for the model
described around Eqs.\ (\ref{OL}) and (\ref{DOS})
where $\hbar\omega_{H_{c2}^{\rm quasi}}/k_{\rm B}T_{c} \!\sim\! 1$ 
and $N_{\rm F}\!\sim \! 30$ at $B\!=\! H_{c2}^{\rm quasi}(0)$.

Figure \ref{fig:delta-self-nonquadra} shows the field-dependence of the expansion
coefficients $\tilde{\Delta}^{(N_{\rm c})}$
calculated self-consistently for
the $s$-wave hexagonal lattice (first column) and the $d$-wave square lattice (second column). 
Singular oscillatory behaviors
are manifest in both cases as in the case of the quadratic dispersion,
which originate from the singular density of states of 
Landau levels.\cite{Rasolt92}
For example, the dominant $\tilde{\Delta}^{(0)}$ component
have a nonzero value from above $H_{c2}^{\rm quasi}$ and
deviates substantially from the quasiclassical square-root behavior
(dotted lines).

\begin{figure}[t]
\includegraphics[width=0.99\linewidth]{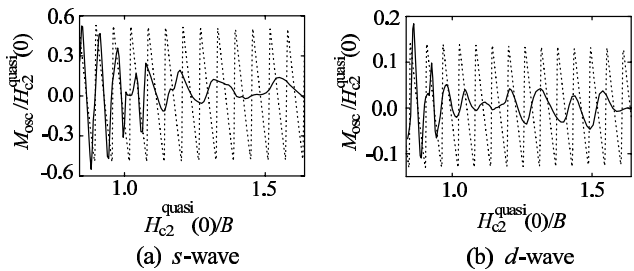}
\caption{Oscillatory part of magnetization $M_{\rm osc}$ for (a) the $s$-wave
and (b) the $d$-wave, 
over $0.8\!\alt\!H_{c2}^{\rm quasi}(0)/B\!\leq \! 1.7$,
i.e.\ $1.25 \!\agt\! B/H_{c2}^{\rm quasi}(0)\!\geq \! 0.59$.
Here  $T\!=\!0.1T_{c}$, 
and a non-quadratic dispersion given by Eq.\ (\ref{DOS}) is used.
The dotted lines are the curves of the corresponding normal state.}
\label{fig:Mosc-self-nonquadra}
\end{figure}

Figure \ref{fig:Mosc-self-nonquadra} displays the corresponding 
oscillatory part of the magnetization $M_{\rm osc}$
calculated numerically by Eq.\ (\ref{mag}), where
normal-state results (dotted lines) are also plotted
for comparison. 
The main features are summarized as follows:
(i) The oscillations are seen to decrease from above the
quasiclassical $H_{c2}^{\rm quasi}$, due to the reentrant behavior
of $\tilde{\Delta}^{(N_{\rm c})}$,
to be reduced considerably around 
$B\!\sim\!0.8H_{c2}^{\rm quasi}$, i.e.\ $H_{c2}^{\rm quasi}/B\! \sim \! 1.25$.
However, they do not disappear completely in lower fields.
(ii) This extra attenuation is due to the broadening of the Landau levels
caused by the pair potential, as in the case of the quadratic dispersion.
Indeed, we have obtained quasiparticle spectra similar to those of
Fig.\  \ref{fig:disp-self}.
(iii) The period of the oscillation remains
unchanged above $\sim\!0.8H_{c2}^{\rm quasi}$, but some irregularity
appears in lower fields.
These features are in agreement with the results by
Norman {\em et al}.\cite{NMA95}
(iv) Little difference can be seen between the $s$- and $d$-wave attenuations.

\subsection{Summary of Two-Dimensional Calculations}
\label{2DSummary}

\begin{figure}[t]
\includegraphics[width=0.99\linewidth]{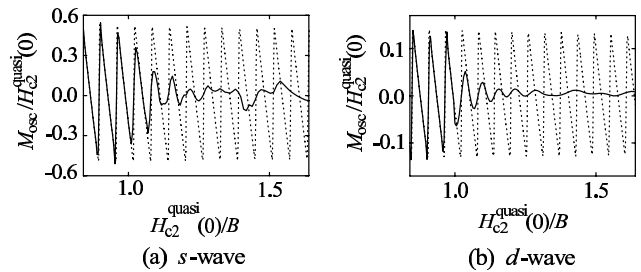}
\caption{Oscillatory part of magnetization $M_{\rm osc}$ for (a) the $s$-wave
and (b) the $d$-wave.
The difference from Fig.\ \ref{fig:Mosc-nonquadra} lies in the use of
quasiclassical $\Delta^{(0)}$'s given by the dotted lines in 
Fig.\ \ref{fig:delta-self-nonquadra}.
}
\label{fig:Mosc-nonquadra}
\end{figure}

Let us summarize results and conclusions from our two-dimensional calculations.
(i) Combining Figs.\ \ref{fig:disp-self} and \ref{fig:Mosc-self}, 
we are now able to attribute the origin of the extra dHvA oscillation
damping unambiguously to the Landau-level broadening 
due to the scattering by the pair potential.
(ii) As may be realized from Fig.\ \ref{fig:Mosc-self-nonquadra},
presence of point nodes along the extremal orbit does not
weaken the attenuation, 
contrary to the statement by Miyake.\cite{Miyake93}
This fact suggests that the attenuation is determined 
by the average gap along the extremal orbit.
(iii) The mean-field theory predicts that the dHvA oscillation comes together
with the oscillatory behaviors in $H_{c2}$ and 
${\tilde\Delta}^{(N_{\rm c})}$. 
This will be so in three dimensional models where
$H_{c2}(T)$ also shows an oscillatory behavior.\cite{Rasolt92}
However, such singular behaviors of $H_{c2}$ have never been identified definitely
in any materials displaying the dHvA oscillation, and
reported $H_{c2}$ curves show more or less
the smooth quasiclassical behavior.
This discrepancy between
the mean-field theory and the dHvA experiments
remains a puzzle to be resolved in the future.
(iv) The oscillation attenuates considerably
around $B\!\sim\!0.8H_{c2}^{\rm quasi}$, i.e.\ $H_{c2}^{\rm quasi}/B\! \sim \! 1.25$, 
although we have set
$\hbar\omega_{H_{c2}^{\rm quasi}}/k_{\rm B}T_{c} \!\sim\! 1$
and $N_{\rm F}\!\gg\! 1$ at $H_{c2}^{\rm quasi}$.
Thus, the two dimensional models fail to explain the experiment by
Terashima {\em et al}.\ \cite{Terashima95} which shows a persistent oscillation down
to $0.2H_{c2}$. In addition, the models cannot say anything about whether 
presence of a line node along the extremal orbit weakens the attenuation.
(v) The approximation of retaining only 
${\tilde\Delta}^{(0)}$ works excellently for 
calculating $M_{\rm osc}$. 
Indeed, we have checked that 
the curves of $M_{\rm osc}$ thereby obtained are almost indistinguishable
from those of Fig.\ \ref{fig:Mosc-self-nonquadra}.
(vi) The discrepancy mentioned in (iii) above suggests that we should rather
use $\tilde{\Delta}^{(0)}$ obtained quasiclassically
to reproduce the smooth behaviors of $H_{c2}$ in real materials.
Figure \ref{fig:Mosc-nonquadra} plots curves of 
$M_{\rm osc}$ calculated using quasiclassical 
$\tilde{\Delta}^{(0)}$, i.e.\ the dotted lines of Fig.\ \ref{fig:delta-self-nonquadra}.
The oscillations are seen more regular than those of Fig.\ \ref{fig:Mosc-self-nonquadra},
but the amplitudes attenuate almost similarly and are reduced considerably around 
$H_{c2}^{\rm quasi}/B\! \sim \! 1.25$.
We hence realize that using the quasiclassical $\tilde{\Delta}^{(0)}$
suffices for the theory of the oscillation damping.
This statement is especially true in the low-field region $\sim\!0.2H_{c2}^{\rm quasi}$
where $\tilde{\Delta}^{(0)}$ approaches to the quasiclassical behavior,
as may be realized from Fig.\ \ref{fig:delta-self-nonquadra}.

\section{\label{sec:3D}Three-Dimensional calculations}

Having clarified basic features of the dHvA oscillation for
two-dimensional models as well as 
the mechanism of the extra oscillation damping,
we proceed to consider three-dimensional models with various gap structures
which are more relevant to real materials.
Our purposes in this section are summarized as follows:
(i) to clarify the connection between the extra dHvA oscillation damping and 
the gap anisotropy by numerical calculations;
(ii) to obtain an analytic formula for the extra oscillation damping;
(iii) to estimate the gap magnitudes of various materials using
the obtained analytic formula.

\subsection{Model}

The one-particle Hamiltonian (\ref{calH}) yields a spherical 
Fermi surface in the normal state.
As for the pairing interaction, we consider three different 
models:
\begin{eqnarray}
V_{{\bf p}{\bf p}'}=
\left\{ 
\begin{array}{l}
\vspace{2mm} \! V_{0}\, W(\xi)W(\xi')  \\ 
\vspace{2mm} \!V_{2}\, W(\xi)(\hat{\bf p}_{x}^{2}\!-\!
\hat{\bf p}_{y}^{2})\, W(\xi') (\hat{\bf p}_{x}^{\,\prime 2}\!-\!
\hat{\bf p}_{y}^{\,\prime 2}) \\
V_{1}\, W(\xi)\,\hat{\bf p}\cdot\hat{\bf c} \,
W(\xi') \,\hat{\bf p}'\cdot\hat{\bf c} \,
\end{array}
\right. .
\label{pVp'3}
\end{eqnarray}
Here $W(\xi)$ is a cut-off function given by Eq.\ (\ref{Wcut}),
$\hat{\bf p}\!\equiv\!(\sin\theta_{\bf p}\cos\varphi_{\bf p},
\sin\theta_{\bf p}\sin\varphi_{\bf p},\cos\theta_{\bf p})$
specifies a point on the Fermi surface,
and $\hat{\bf c}\!\equiv\!(\sin\theta_{\bf c}, 0,\cos\theta_{\bf c})$
denotes the direction of the crystal $c$-axis,
in the coordinate frame where
${\bf B}\!\parallel\!\hat{\bf z}$.
Again the latter two models of Eq.\ (\ref{pVp'3}) 
are beyond the original spherical interaction (\ref{pVp'}),
but they are convenient for the above-mentioned purposes.
In zero field, Eq.\ (\ref{pVp'3}) yield
\begin{eqnarray}
{\underline \Delta}_{{\bf p}}=
\left\{ 
\begin{array}{ll}
\vspace{2mm} \! \Delta_{0} W(\xi)\,i {\underline \sigma}_{2} & \mbox{: $s$-wave} \\ 
\vspace{2mm} \!\Delta_{0} W(\xi)(\hat{\bf p}_{x}^{2}\!-\!
\hat{\bf p}_{y}^{2})\,i{\underline \sigma}_{2} & \mbox{: $d_{x^2-y^2}$-wave}\\
\!\Delta_{0} W(\xi)\,\hat{\bf p}\cdot\hat{\bf c} \,\,
i{\underline \sigma}_{3}{\underline \sigma}_{2} & \mbox{: $p_{z}$-wave}
\end{array}
\right. ,
\label{gap3D}
\end{eqnarray}
which denote the isotropic $s$-wave state, 
a three-dimensional $d_{x^2-y^2}$-wave state with four point nodes 
in the extremal orbit perpendicular to ${\bf B}$, 
and the $p$-wave polar state with a line node 
perpendicular to $\hat{\bf c}$, respectively.
The corresponding $\underline{{\bar\Delta}}^{\!(N_{{\rm c}}m)}_{N_{{\rm r}}p_{z}}$
in Eq.\ (\ref{pair2}) can be written as
\begin{subequations}
\label{pair3D}
\begin{equation}
\underline{{\bar\Delta}}^{\!(N_{{\rm c}}m)}_{N_{{\rm r}}p_{z}}
= \left\{ 
\begin{array}{ll}
\vspace{2mm}  \! {\tilde\Delta}^{(N_{\rm c})} W(\xi) \, 
\delta_{m0} \,i {\underline \sigma}_{2} \\ 
\vspace{2mm} \! {\tilde\Delta}^{(N_{\rm c})} W(\xi)
 \sin^{2}\!\theta_{\bf p}\, 
 {\displaystyle\frac{\delta_{m2}\!+\!\delta_{m-2}}{2}}
\,i{\underline \sigma}_{2} \\
\!{\tilde\Delta}^{(N_{\rm c})} W(\xi)
\biggl[\cos\theta_{\bf p}\!\cos\theta_{\bf c}\, \delta_{m0}
\vspace{1mm}\\
\hspace{3mm}+
\sin\theta_{\bf p}\!\sin\theta_{\bf c} \,{\displaystyle\frac{\delta_{m1}\!+\!\delta_{m-1}}{2}}
\biggr]
i{\underline \sigma}_{3}{\underline \sigma}_{2}
\end{array}
\right. ,
\label{pair3D1}
\end{equation}
where $\xi\!\equiv\! \hbar^{2}(N_{\rm r}/l_{B}^{2}\!+\!p_{z}^{2})/2m_{\rm e}
\!-\!\varepsilon_{\rm F}$,
$\theta_{\bf p}\!\equiv\!\tan^{-1}\bigl(\frac{\sqrt{N_{{\rm r}}}/l_{B}}{p_{z}}\bigr)$,
and ${\tilde\Delta}^{(N_{{\rm c}})}$ is defined by
\begin{equation}
{\tilde\Delta}^{(N_{{\rm c}})}
= \left\{
\begin{array}{l}
\vspace{0mm}
\displaystyle
\!\frac{V_{0}}{4\pi l_{\! B}^{2}}\sum_{ N_{{\rm r}}'p_{z}'}\!
W_{\ell}(\xi')\,
{{\bar\Phi}}^{(N_{{\rm c}},0)}_{N_{{\rm r}}'p_{z}'} 
\\
\displaystyle
\!\frac{V_{2}}{4\pi l_{\! B}^{2}}\sum_{ N_{{\rm r}}'p_{z}'}\!
W_{\ell}(\xi') \sin^{2}\!\theta_{{\bf p}'}\,\frac{
{{\bar\Phi}}^{(N_{{\rm c}},2)}_{N_{{\rm r}}'p_{z}'} \!+\! 
{{\bar\Phi}}^{(N_{{\rm c}},-2)}_{N_{{\rm r}}'p_{z}'} }{2}
\vspace{1mm}
\\
\displaystyle
\!\frac{V_{1}}{4\pi l_{\! B}^{2}}\sum_{ N_{{\rm r}}'p_{z}'}\!
W_{\ell}(\xi') \biggl[ \cos\theta_{{\bf p}'}\!\cos\theta_{\bf c}\,
{\bar\Phi}^{(N_{{\rm c}},0)}_{N_{{\rm r}}'p_{z}'}
\vspace{1mm}\\
\displaystyle
\hspace{12mm}+
\sin\theta_{{\bf p}'}\!\sin\theta_{\bf c}
\frac{
{{\bar\Phi}}^{(N_{{\rm c}},1)}_{N_{{\rm r}}'p_{z}'} \!+\! 
{{\bar\Phi}}^{(N_{{\rm c}},-1)}_{N_{{\rm r}}'p_{z}'} }{2}\,
\biggr]
\end{array}
\right.
\, .
\label{pair3D2}
\end{equation}
\end{subequations}
Here we have adopted a normalization for 
${\tilde\Delta}^{(N_{\rm c})}$
different from Eq.\ (\ref{pair3}) so that this 
quantity acquires a direct correspondence to the maximum
gap $\Delta_{0}$ in Eq.\ (\ref{gap3D}).
The factors $\frac{1}{2}$ in the second and third cases stem from 
$\cos2\varphi_{\bf p}$ and $\cos\varphi_{\bf p}$ in Eq.\ (\ref{gap3D}),
respectively.

The coefficients $\Delta^{(N_{\rm c})}\!=\!\Delta^{(N_{\rm c})}(B,T)$ 
in Eq.\ (\ref{pair3D}) completely specify the pair potential,
as already mentioned.
Based on the reasoning given in Sec.\ \ref{2DSummary}(vi),
we here adopt a quasiclassical
$\tilde{\Delta}^{(N_{\rm c})}$
rather than the fully self-consistent one.
Then the dominant ${\tilde\Delta}^{(0)}$
near $H_{c2}$ follows the mean-field square-root behavior
to an excellent approximation:
\begin{eqnarray}
{\tilde\Delta}^{(0)}=a(1-B/H_{c2})^{1/2} \, .
\label{Del(0)}
\end{eqnarray}
See the dotted lines in Figs.\ \ref{fig:delta-self} and \ref{fig:delta-self-nonquadra},
for example.
In addition, other components ${\tilde\Delta}^{(N_{\rm c}>0)} $ can be neglected 
for the relevant region $B\!\agt\! 0.1H_{c2}$, as pointed out in Sec.\ \ref{2DSummary}(v).
We hence use the lowest-Landau-level approximation of retaining only ${\tilde\Delta}^{(0)}$.
The coefficient $a\!=\! a(T)$ 
in Eq.\ (\ref{Del(0)}) is determined by requiring that the maximum of
\begin{eqnarray}
\frac{1}{{\cal V}}\!\int \! d{\bf R}\, \left| \int \!
d{\bf r}\,\Delta({\bf r}_{1},{\bf r}_{2})\,
{\rm e}^{-i{\bf p}\cdot{\bf r}/\hbar}\, \right|^{2}
\label{DeltaAverage}
\end{eqnarray}
be equal to $\Delta_{0}^{2}(1\!-\! B/H_{c2})$,
where $\Delta_{0}(T)$ denotes the maximum gap obtained from the weak-coupling theory.
This procedure yields
\begin{eqnarray}
a\approx\sqrt{0.5} \, ,
\label{a^2=0.5}
\end{eqnarray}
for all the three cases of Eq.\ (\ref{pair3D}).
Substituting Eq.\ (\ref{Del(0)}) into Eq.\ (\ref{D-Dt1}) with
the choice $\hbar\omega_{\rm D}\!\sim\! 10 \Delta_{0}(T\!=\!0)$,
the off-diagonal elements of Eq.\ (\ref{BdG2}) are fixed completely.

The above non-self-consistent procedure has another advantage
that we can choose $\hbar\omega_{B= H_{c2}}$ and
$\varepsilon_{\rm F}$ in Eq.\ (\ref{calH2}) independently.
We have set 
\begin{equation}
\hbar\omega_{H_{{\rm c}2}}= k_{\rm B}T_{c} \hspace{5mm} \mbox{at } T=0 \, ,
\end{equation}
in accordance with $\hbar\omega_{H_{{\rm c}2}}/k_{\rm B}T_{c}\!=\!1\!\sim\!3$
and $N_{\rm F}\!\gg\! 1$ for relevant materials (see Table \ref{table1} below).
Also, we have chosen $\varepsilon_{\rm F}$ in such a way that 
there are about $50$ Landau levels below $\varepsilon_{\rm F}$ 
for the extremal orbit at $H_{c2}$.
Now, the matrix elements of Eq.\ (\ref{BdG2}) are specified completely.
Hexagonal, square, and hexagonal lattices are assumed for the
three cases of Eq.\ (\ref{pair3D}), respectively.

\subsection{Numerical Procedures}
\label{NumProc3D}

The wavevector $p_{z}$ of $0 \!\leq\! p_{z} \!\leq\! 1.2p_{\rm F}$ is discretized 
into $\sim\! 1000$ points with an equal interval.
For each of them, we have diagonalized Eq.\ (\ref{BdG2})
with the same procedure as described in Sec.\ \ref{NumProc2D}.
The obtained results are substituted
into Eq.\ (\ref{Omega}) to calculate the magnetization by Eq.\ (\ref{mag}).
All the calculations are performed at $T\!=\!0$.

\begin{figure}[t]
\includegraphics[width=0.99\linewidth]{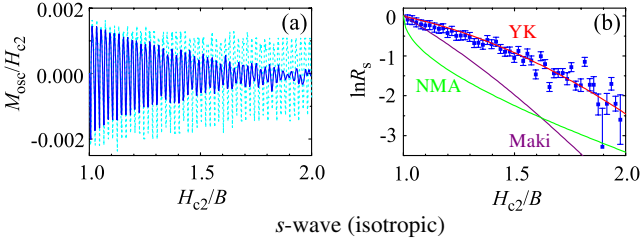}
\caption{(a) The oscillatory part
of the magnetization $M_{\rm osc}$ in the vortex state (blue line)
as compared with the normal-state one (sky-blue line)
for the $s$-wave model of Eq.\ (\ref{gap3D}) at $T\!=\!0$.  
(b) The corresponding Dingle plot (points with error bars)
as compared with various theoretical predictions; see text for details.}
\label{fig:dHvA-s}
\end{figure}

\subsection{Numerical Results}
\label{NumResults3D}

We first focus on the dHvA oscillation of the
$s$-wave model in Eq.\ (\ref{gap3D})
to clarify its basic features.
Using our numerical data, 
we also test the applicability of various theoretical formulas
presented so far.

\begin{figure}[t]
\includegraphics[width=0.45\linewidth]{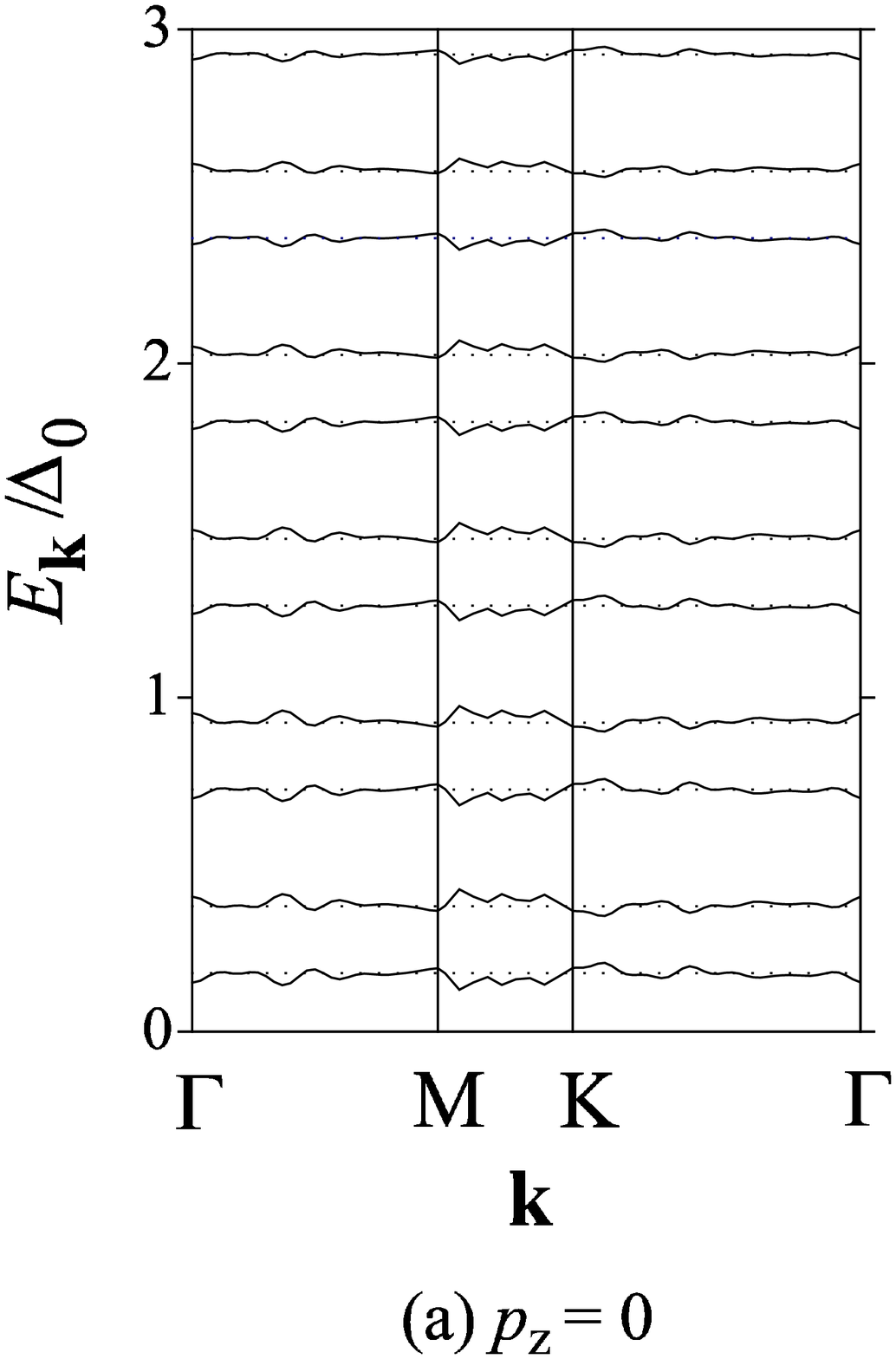}\hspace{2mm}
\includegraphics[width=0.45\linewidth]{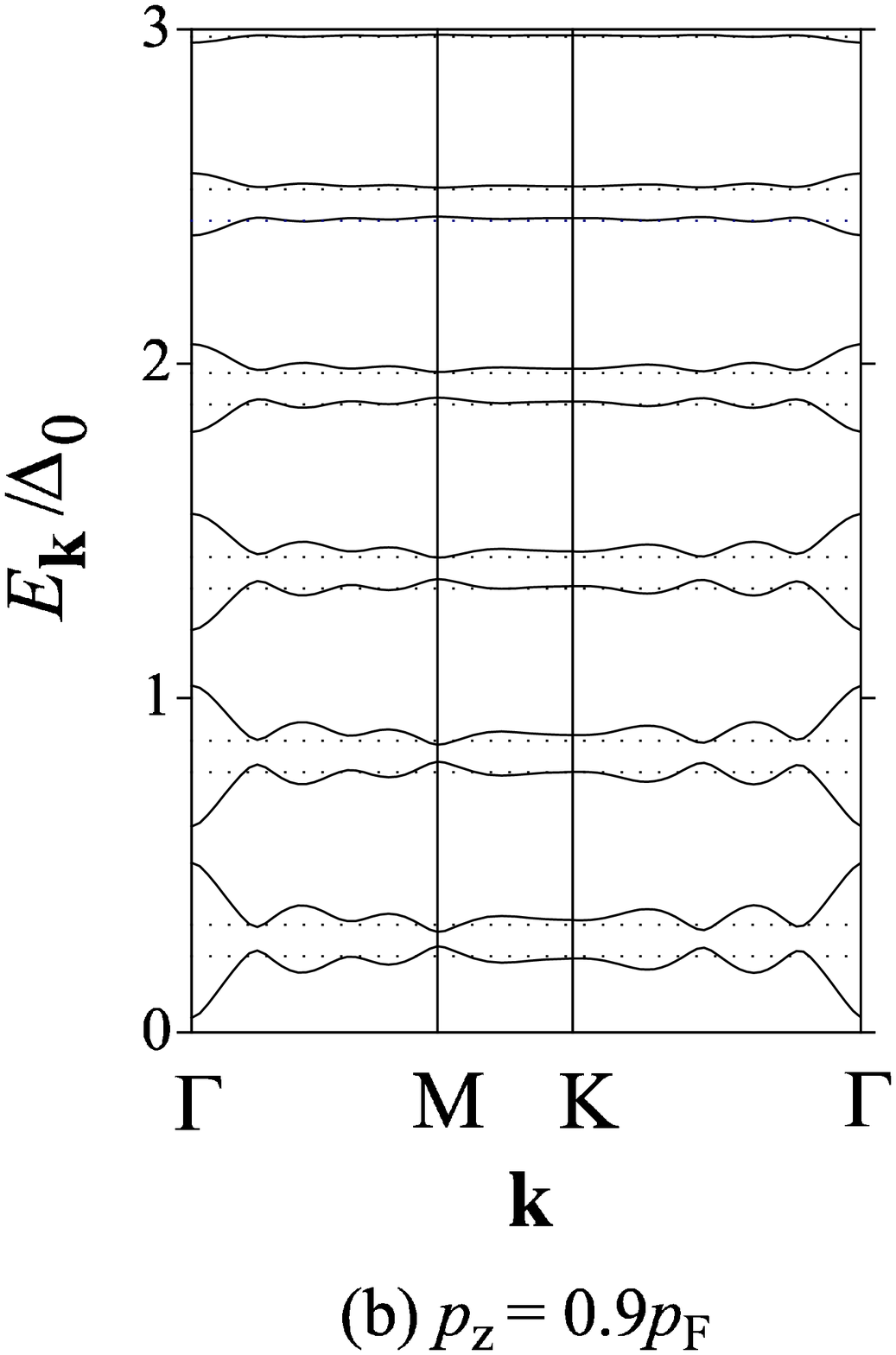}\\
\caption{Quasiparticle dispersion in the magnetic Brillouin zone
for the $s$-wave model at $B\!=\! 0.968H_{c2}$.
(a) $p_z$=0; (b) $p_z$=0.9$p_{\rm F}$.}
\label{fig:disp-pz}
\end{figure}

Figure \ref{fig:dHvA-s}(a)
presents oscillation of the $s$-wave magnetization (blue line)
as compared with the normal-state one (sky-blue line).
With $\hbar\omega_{H_{c2}}\!=\!k_{\rm B}T_{c}$,
the oscillation is observed to persist down to a fairly low field of
$H_{c2}/B \!\alt\! 1.8$, i.e., $B\!\agt
\! 0.55H_{c2}$,
which is smaller than $0.8H_{c2}$
around which $\hbar\omega_{B}$ becomes equal to
the spatial average of the energy gap, Eq.\ (\ref{DeltaAverage}).
This is partly because the gap is smaller within the extremal orbit,
as shown quasiclassically by Brandt {\em et al}.\cite{Brandt67}
Indeed, Fig.\ \ref{fig:disp-pz} calculated at $B\!=\! 0.968H_{c2}$
demonstrates that the dispersion for 
$p_{z}\!=\!0$ is smaller than that for $p_{z}\!=\!0.9p_{\rm F}$.
This tendency remains in the high-field region of $B\!\agt\!0.5H_{c2}$.

It has become conventional to express this extra attenuation in the vortex state
by introducing an additional factor $R_{\rm s}$ for the dHvA oscillation amplitude:
\begin{equation}
R_{\rm s}=\exp\!\left(-\frac{\pi}{\omega_{B}\tau_{\rm s}}\right)
=\exp\!\left(-\frac{2\pi^{2}k_{\rm B}T_{\Delta}}{\hbar\omega_{B}}\right) \, ,
\end{equation}
where the parameters $\tau_{\rm s}$ and $T_{\Delta}$ are directly connected with the extra
Landau-level broadening $\Gamma_{\rm s}$ in the vortex state as $\Gamma_{\rm s}\!=\!
\hbar/2\tau_{\rm s}\!=\! \pi k_{\rm B}T_{\Delta}$.
The points with error bars in Fig.\ \ref{fig:dHvA-s}(b) 
shows $\ln R_{\rm s}$ as a function of $1/B$, i.e.\
the Dingle plot,
obtained by numerical differentiation.
This extra damping at high fields shows the behavior
$\propto\!1\!-\!B/H_{c2}$ in the logarithmic scale,
but irregularity sets in around $0.55H_{c2}$
where the oscillation disappears.
We attribute this irregularity to the effect of the bound-state formation
in the core region.

The lines in Fig.\ \ref{fig:dHvA-s}(b)
are the predictions from various theoretical formulas.
Maki's formula\cite{Maki91} reproduces the correct functional behavior
$\propto\!1\!-\!B/H_{c2}$ at high fields,
but the prefactor is seen too large.
The NMA formula,\cite{NMA95}
deduced from the two-dimensional self-consistent numerical results 
with $N_{\rm F}\!\sim\!10$ at $H_{c2}$,
predicts a more rapid attenuation incompatible with our numerical data.
One reason for this discrepancy
may originate from the fact that their numerical data 
with $N_{\rm F}\!\sim\! 10$ at $H_{c2}$
are not appropriate for obtaining an analytic formula by fitting.
Another may be attributed to the difference in dimensions.
Indeed, the dHvA oscillation in three dimensions 
differs from that in two dimensions 
on the point that some finite region 
$\delta p_{z}$ around the extremal orbit is relevant.
Most of the Landau levels in the region 
do not satisfy the particle-hole
symmetry with respect to $\varepsilon_{\rm F}$, 
so that the effect of the pair potential becomes smaller
than that in two dimensions.
Another theory by Dukan and Te\v sanovi\'c,\cite{Dukan95} 
which would predict $R_{\rm s}\!=\!0$ in the clean limit of $T\!=\!0$, 
is also inconsistent with the data.

The red line in Fig.\ \ref{fig:dHvA-s}(b) 
is due to our formula for the extra Dingle temperature:
\begin{eqnarray}
k_{\rm B}T_{\Delta}=
0.5\hspace{0.3mm}\tilde{\Gamma}\hspace{0.3mm}\langle|\Delta_{\bf p}|^{2}\rangle_{\rm eo}
\hspace{0.3mm}\frac{m_{\rm b}c}{\pi e\hbar}
\, \frac{1\!-\!B/H_{c2}}{B} \, ,
\label{T_D}
\end{eqnarray}
which is derived in Appendix \ref{App:dHvA-analytic}
based on the second-order perturbation with respect to the pair potential.
Here $\langle|\Delta_{\bf p}|^{2}\rangle_{\rm eo}$ denotes the average gap 
along the extremal orbit at $B \!=\! 0$, and 
$m_{\rm b}$ is the band mass.
The numerical constant $0.5$ stems from Eq.\ (\ref{a^2=0.5}),
and $\tilde{\Gamma}$ is a dimensionless quantity 
characterizing the Landau-level broadening due to the pair potential.
This unknown parameter $\tilde{\Gamma}$ is determined
by a best fit to the $s$-wave numerical data, i.e.\ the points with
error bars in Fig.\ \ref{fig:dHvA-s}(b).
This procedure yields
\begin{eqnarray*}
\tilde{\Gamma}\!=\!0.125 \, .
\end{eqnarray*}
We observe in  Fig.\ \ref{fig:dHvA-s}(b) that Eq.\ (\ref{T_D}), 
which predicts the dependence $\propto\!1\!-\!B/H_{c2}$ for $\ln R_{\rm s}$,
agrees with the numerical results.
This formula will be seen below to 
reproduce other numerical data excellently 
without any adjustable parameters.

\begin{figure}[t]
\includegraphics[width=0.99\linewidth]{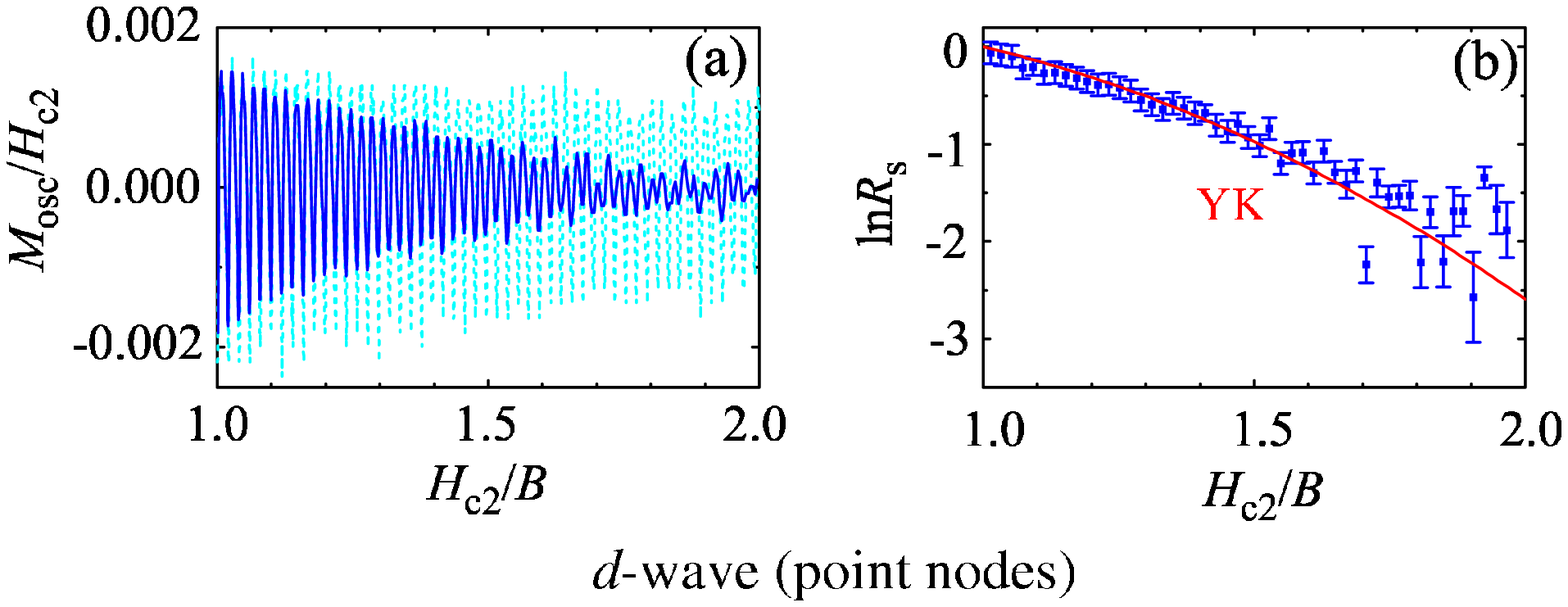}
\caption{(a) The oscillatory part
of the magnetization $M_{\rm osc}$ in the vortex state (blue line)
as compared with the normal-state one (sky-blue line)
for the $d$-wave model of Eq.\ (\ref{gap3D}) at $T\!=\!0$.  
(b) The corresponding Dingle plot (points with error bars)
as compared with the theoretical prediction (\ref{T_D}).}
\label{fig:dHvA-d}
\end{figure}

A difference of Eq.\ (\ref{T_D}) 
from Maki's formula\cite{Maki91} lies in the prefactor
where the Fermi velocity $v_{\rm F}$ is absent.
Indeed, a dimensional analysis on the second-order perturbation
tells us that the Landau-level broadening in the vortex state
should be of order
$|\tilde{\Delta}^{\!(0)\!}(B)|^{2}/\hbar \omega_{B}$,
where $\tilde{\Delta}^{\!(0)\!}(B)\!\propto\! \sqrt{\langle|\Delta_{\bf p}|^{2}\rangle_{\rm eo}
(1\!-\! B/H_{c2})}$ is essentially the average gap along the extremal orbit.
This leads to Eq.\ (\ref{T_D}) except for the numerical constant.

We now turn our attention to see how the presence of point nodes
affect the dHvA oscillation.
Figure \ref{fig:dHvA-d}(a) shows the oscillation 
of the $d$-wave magnetization (blue line)
as compared with the normal-state one (sky-blue line). 
Although the $d$-wave gap in Eq.\ (\ref{gap3D})
has four point nodes on the Fermi surface along the extremal orbit,
the damping is seen strong and not much different from the $s$-wave case.
From this fact, we may conclude that it is the
average gap along the extremal orbit which is relevant 
for the extra dHvA oscillation damping.
Figure \ref{fig:dHvA-d}(b) presents the corresponding Dingle plot
(points with error bars), which is compared with the prediction of Eq.\ (\ref{T_D}).
The formula with the average gap $\langle|\Delta_{\bf p}|^{2}\rangle_{\rm eo}$ 
reproduces the numerical result 
for $H_{c2}/B \!\alt \! 1.8$ excellently without adjustable parameters, 
thereby providing a strong support for the above statement.

This $d$-wave result is in disagreement with
Miyake's theory that point nodes in the extremal orbit 
should weaken the attenuation.\cite{Miyake93}
Indeed, Miyake's theory is based on a
semiclassical quantization
for the expression of the electron number $N_{e}$ at $B\!=\!0$.
Neither his starting point $N_{e}(B\!=\!0)$ nor the use of the semiclassical quantization
may be justified for describing the dHvA oscillation observed
mainly near $H_{c2}$.

\begin{figure}[t]
\includegraphics[width=0.99\linewidth]{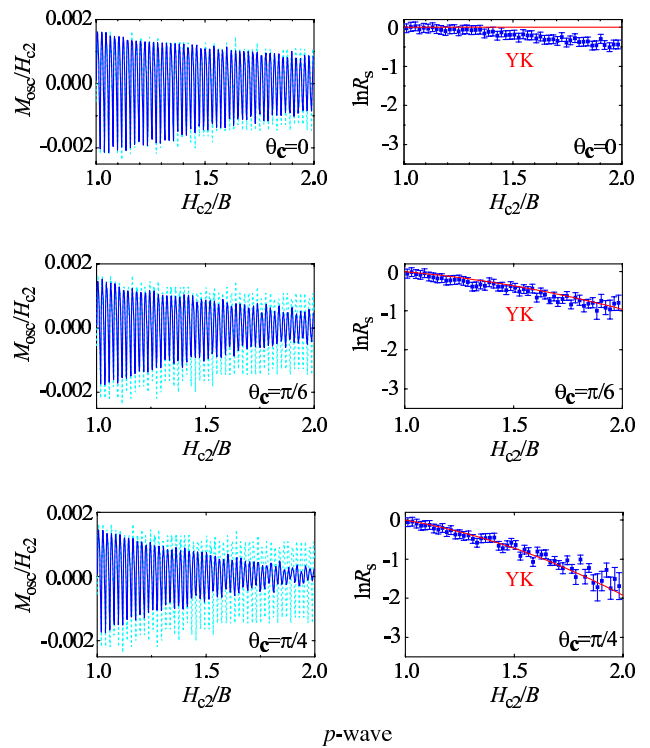}
\caption{Left figures: the oscillatory part
of the magnetization $M_{\rm osc}$ in the vortex state (blue lines)
as compared with the normal-state one (sky-blue lines)
for the $p$-wave model of Eq.\ (\ref{gap3D}) at $T\!=\!0$. 
The crystal $c$ axis, which is perpendicular to the
nodal plane, is tilted from the field ${\bf B}$
by $\theta_{\bf c}\!=\! 0$ (top), 
$\theta_{\bf c}\!=\! \pi/6$ (second),
and $\theta_{\bf c}\!=\! \pi/4$ (bottom).
Right figures: the corresponding Dingle plots (points with error bars)
compared with Eq.\ (\ref{T_D}).}
\label{fig:dHvA-p}
\end{figure}
\begin{table*}[t]
\caption{\label{table1}
Parameters characterizing three superconductors displaying the dHvA oscillation
in the vortex state, 
together with values of the average gap along the extremal orbit 
$\sqrt{\langle|\Delta_{\rm p}|^2\rangle_{\rm eo}}$ estimated by Eq.\ (\ref{T_D}).
Here the symbol $\alpha$ and $\gamma$ are band indices.
These values of $\sqrt{\langle|\Delta_{\rm p}|^2\rangle_{\rm eo}}$
are to be compared with the band- and/or 
angle-averaged quantity $\Delta(0)$
extracted from a specific-heat experiment,\cite{Movshovich94} 
a far-infrared measurement,\cite{Clayman71} a tunneling experiment,\cite{Lee75} 
or a Raman-scattering experiment.\cite{Hackl83}}
\begin{ruledtabular}
\begin{tabular}{cccccccc}
Compound & $T_{c}$ (K) & $H_{c2}(T\! =\! 0)$ (T) & $m_{\rm b}/m_{e}$ & 
$\hbar\omega_{H_{c2}}$ (K) & $\hbar\omega_{H_{c2}}/k_{\rm B}T_{c}$
&  $\sqrt{\langle|\Delta_{\rm p}|^2\rangle_{\rm eo}}$ (meV)  & $\Delta(0) $ (meV)
\\
\hline
NbSe$_{2}$  & 7.2\cite{Janssen98} & 8.01 ($\theta=68.6^{\circ}$)
\cite{Janssen98} & 0.61 ($\alpha$)\cite{Janssen98} & 17.6\cite{Janssen98} & 2.20\cite{Janssen98}
& 1.1 ($\pm 0.04$)   & 1.1 \cite{Clayman71,Lee75}
\\ 
Nb$_{3}$Sn & 18.3 \cite{Harrison94} & 19.7 ($H\!\parallel\! c$) \cite{Harrison94}
 & 1.10 ($\gamma$) \cite{Harrison94} & 24.1 \cite{Harrison94} & 1.31\cite{Harrison94}
&  3.2 ($\pm 0.19$)   & 3.2 \cite{Hackl83} \\ 
YNi$_{2}$B$_{2}$C & 14.5 \cite{Terashima95} & 8.8 ($H\!\parallel\! c$) \cite{Terashima95}
 & 0.35 ($\alpha$) \cite{Terashima95} &
33.8 \cite{Terashima95} & 2.33  \cite{Terashima95}
&  1.5 ($\pm 0.28$)  & 2.5 \cite{Movshovich94}
\\ 
\end{tabular}
\end{ruledtabular}
\end{table*}

We finally consider the $p$-wave model with a line node
in Eq.\ (\ref{gap3D}) to double-check 
the applicability of Eq.\ (\ref{T_D}).
The left figures in Fig.\ \ref{fig:dHvA-p} display the 
dHvA oscillation for the line-node model,
where the crystal $c$-axis is tilted  from the magnetic-field direction 
by $\theta_{\bf c}\!=\! 0$ (top), 
$\theta_{\bf c}\!=\! \pi/6$ (second),
and $\theta_{\bf c}\!=\! \pi/4$ (bottom).
The damping is seen weakest in the top figure
where the gap vanishes along the extremal orbit,
but increases gradually as finite gap opens along the orbit 
for $\theta_{\bf c}\!=\! 0\!\rightarrow\! \pi/4$.
These results indicate conclusively that the average gap along the extremal
orbit is relevant for the extra dHvA attenuation.
However, the non-zero extra damping in the top figure implies that
not only the extremal orbit alone but 
some finite region around it contributes to the extra damping.
Theoretically, this corresponds to the fact that
we have to perform the Fresnel 
integral $\int_{-\infty}^{\infty}\!
\exp[-i(\sqrt{2\pi N_{\rm F}}\,p_{z}/p_{\rm F})^{2}]\, dp_{z}$
for obtaining the LK formula in the normal state.
Our data show that this off-extremal-orbit contribution cannot 
be neglected in the case where the gap vanishes exactly and completely
at the extremal orbit.
However, this off-extremal contribution is expected to become less important 
where finite gap is present along the extremal orbit.
The right figures in Fig.\ \ref{fig:dHvA-p} show the corresponding Dingle plot
(points with error bars), which is compared with the prediction of Eq.\ (\ref{T_D}).
Except for the weak damping of $\theta_{\bf c}\!=\!0$ due to
the off-extremal-orbit contribution, the formula is observed to 
reproduce the numerical results excellently.

\section{\label{sec:estimation}
Estimation of Energy Gap}

Our calculations in Sec.\ \ref{NumResults3D} have clarified that 
(i) the gap anisotropy can be detected
by measuring the extra dHvA oscillation damping in the vortex state, 
and (ii) Eq.\ (\ref{T_D}) is particularly useful for this purpose.
Using the formula,
we finally provide quantitative estimations of the average gap 
along the extremal orbit for several 
materials displaying the dHvA oscillation in the vortex state. 
Table \ref{table1} summarizes parameters describing
three relevant materials.
These materials commonly have fairly high $T_{c}$'s, and the ratio 
$\hbar\omega_{H_{c2}}/k_{\rm B}T_{c}$ ranges from $1$ to $3$.
These features seem to be basic conditions for observing
the dHvA oscillation in the vortex state. 
The values for $\sqrt{\langle|\Delta_{\rm p}|^2\rangle_{\rm eo}}$ are obtained
by applying Eq.\ (\ref{T_D}) to the observed dHvA attenuation
in the vortex state. 
In doing so, we have adopted as $m_{b}$ in Eq.\ (\ref{T_D})
the values from dHvA experiments
rather than those from band calculations,
as indicated by the theory of Luttinger.\cite{Luttinger61}
For comparison, we have also listed the values $\Delta(0)$
estimated by a specific-heat experiment,\cite{Movshovich94} 
a far-infrared measurement,\cite{Clayman71} a tunneling experiment,\cite{Lee75} 
or a Raman-scattering experiment.\cite{Hackl83}
Thus, $\Delta(0)$ is expected to represent band- and/or angle-averaged
energy gap.
As seen in Table \ref{table1}, the two quantities coincide excellently for NbSe$_{2}$ and
Nb$_{3}$Sn, indicating uniformly opened gap in these materials.
On the other hand, $\sqrt{\langle|\Delta_{\rm p}|^2\rangle_{\rm eo}}\!=\! 1.5$ 
for the $\alpha$ band of YNi$_{2}$B$_{2}$C is considerably smaller 
than $\Delta(0)\!=\! 2.5$ from a specific-heat experiment.\cite{Movshovich94}
This fact implies that YNi$_{2}$B$_{2}$C have large 
band- and/or angle-dependent gap anisotropy.
Indeed, Bintley {\em et al}.\ \cite{Meeson02} have recently carried out 
a detailed dHvA experiment on this material,
rotating the field direction and observing the extra attenuation.
They have reported a large angle dependence of the attenuation magnitude.
They have also pointed out that their result is in agreement with the model
with point nodes presented by Izawa {\em et al}.\ \cite{Izawa02}
based on a thermal-conductivity measurement.

\section{\label{sec:summary}
Summary}

We have carried out the first three-dimensional numerical calculations
on the dHvA oscillation in the vortex state
for various gap structures.
We have thereby clarified the relation between
gap anisotropy and persistence of the oscillation.
We have also derived an analytic formula for the extra dHvA attenuation
in the vortex state.

Our main results are given by Figs.\ \ref{fig:dHvA-s}-\ref{fig:dHvA-p} and Eq.\ (\ref{T_D}).
Those figures indicate clearly that the extra dHvA attenuation in the vortex state
is directly connected with the average gap along the extremal orbit at $B\!=\! 0$.
The derived formula (\ref{T_D}) have been shown to 
reproduce the numerical results excellently.
Our theory attributes the origin of the extra dHvA damping
to the Landau-level broadening caused by the pair potential.
Hence the periodicity of the vortex lattice assumed here 
is almost irrelevant,
and the theory is applicable also to the cases
with irregularity such as a random array of vortices.
Using Eq.\ (\ref{T_D}), we have estimated average gap amplitudes along the extremal orbit
for NbSe$_{2}$, Nb$_{3}$Sn, and YNi$_{2}$B$_{2}$C.
The results indicate presence of large gap anisotropy in YNi$_{2}$B$_{2}$C.

Thus, we have shown explicitly that the dHvA effect
in the vortex state can be a powerful tool to probe
the average gap along the extremal orbit.
Our results imply that,
by rotating the field direction
and observing the attenuation amplitude,
we can obtain unique information on the band- and/or angle-dependent
gap structure.
Such an experiment has recently been performed on UPd$_{2}$Al$_{3}$
by Inada {\em et al}.\ \cite{Inada99} and on  YNi$_{2}$B$_{2}$C
by Bintley {\em et al}.,\ \cite{Meeson02} and
the latter group indeed has detected large gap anisotropy
in the $ab$ plane.
Equation (\ref{T_D}) will be useful in similar experiments
for estimating band- and/or angle-dependent gap amplitudes.

\begin{acknowledgments}
We are grateful to F.\ J.\ Ohkawa for enlightening discussions.
This research is supported by Grant-in-Aid for Scientific Research 
from the Ministry of Education, Culture, Sports, Science, and Technology
of Japan.
\end{acknowledgments}

\appendix

\section{\label{App:Thermo}Thermodynamic Potential}

The Luttinger-Ward thermodynamic potential corresponding to Eq.\ (\ref{BdG})
is given by \cite{Kita96}
\begin{eqnarray}
&& \Omega= -\frac{k_{\rm B}T}{2} \sum_{n} {\rm Tr}\ln \!\left[ 
\begin{array}{cc}
\vspace{1mm}
{\cal H}\!-\!i\varepsilon_{n}& 
\Delta \\ 
\Delta^{\! \dagger}& 
-{\cal H}^{*}\!-\!i\varepsilon_{n}
\end{array}
\right]
\nonumber \\ 
&& \hspace{8mm}\times\left[ 
\begin{array}{cc}
\vspace{1mm}
{\rm e}^{i\varepsilon_{n}0_{+}} & 0 \\ 
0 & 
{\rm e}^{-i\varepsilon_{n}0_{+}}
\end{array}
\right]-\frac{1}{2}
{\rm Tr}\,\Delta^{\!\dagger}
\Phi\, ,
\label{Omega1}
\end{eqnarray}
where we have adopted a compact notation of using $x\!\equiv\!{\bf r}\sigma$
with $\Delta_{\sigma_{1}\sigma_{2}}({\bf r}_{1},{\bf r}_{2})\!\rightarrow\!
\Delta(x_{1},x_{2})$, etc.,
and Tr also implies both integration and summation 
over ${\bf r}$ and $\sigma$, respectively.
The quantity $\varepsilon_{n}/\hbar$ denotes the Matsubara frequency,
and $0_{+}$ is an infinitesimal positive constant.
Now, Eq.\  (\ref{BdG}) tells us that
the first matrix in Eq.\ (\ref{Omega1}) can be diagonalized as \cite{Kita96}
\begin{eqnarray}
&& \hspace{-1mm}\left[ 
\begin{array}{cc}
\vspace{1mm}
{\cal H}(x,x')& 
\Delta(x,x') \\ 
\Delta^{\! \dagger}(x,x')& 
-{\cal H}^{*}(x,x')
\end{array}
\right]
\nonumber \\ 
&&\hspace{-4mm}=\!\sum_{s}\!\left[ 
\begin{array}{cc}
\vspace{1mm}
u_{s}^{*}(x)\! & \! 
-v_{s}(x) \\ 
v_{s}^{*}(x)\! & \! 
-u_{s}(x)
\end{array}
\right]\!\!
\left[ 
\begin{array}{cc}
\vspace{1mm}
E_{s}\! & \! 
0 \\ 
0\! & \! 
-E_{s}
\end{array}
\right]\!\!
\left[ 
\begin{array}{cc}
\vspace{1mm}
u_{s}(x')\! & \!
v_{s}(x') \\ 
-v_{s}^{*}(x')\! & \!
-u_{s}^{*}(x')
\end{array}
\right].
\nonumber \\
\label{diagonalize}
\end{eqnarray}
Substituting Eq.\ (\ref{diagonalize}) into Eq.\ (\ref{Omega1}), 
the first term on the right-hand side becomes
\begin{eqnarray}
&& \hspace{-5mm}-\frac{k_{\rm B}T}{2} \sum_{n s} \int \!{\rm d}x \,
\bigl\{\!
\bigl[|u_{s}(x)|^{2}{\rm e}^{z_{n}0_{+}}+
|v_{s}(x)|^{2}{\rm e}^{-z_{n} 0_{+}}\bigr]
\nonumber \\ &&\hspace{20mm}\times\ln(E_{s}\!-\! z_{n})
\nonumber \\
&&+\bigl[|v_{s}(x)|^{2}{\rm e}^{z_{n}0_{+}}+
|u_{s}(x)|^{2}{\rm e}^{-z_{n} 0_{+}}\bigr]\!\ln(-E_{s}\!-\! z_{n})\bigr\},
\nonumber \\
\label{Omega2}
\end{eqnarray}
with $z_{n}\!\equiv \!i\varepsilon_{n}$.
The summation over $n$ are then transformed with a standard technique\cite{LW60}
into a contour integral just above and below the real axis,
using $f(z)\equiv ({\rm e}^{z/k_{\rm B}T}\!+\!1)^{-1}$ and $f(-z)$ for 
the terms with ${\rm e}^{z_{n} 0_{+}}$ and ${\rm e}^{-z_{n} 0_{+}}$, respectively.
Considering the poles inside the two contours and using
$\int [|u_{s}(x)|^{2}\!+\!
|v_{s}(x)|^{2}]{\rm d}x\!=\!1$, we obtain Eq.\ (\ref{Omega}).

\section{\label{App:Basis}Basis Functions and Overlap Integrals}

We here present explicit expressions for the quantities
appearing in Eqs.\ (\ref{Dexp1}) and (\ref{D-Dt});
see Ref.\ \onlinecite{Kita98-2} for their detailed derivations.
It should be noted that we here adopt
the symmetric gauge (\ref{Asymm})
which is more convenient
than the Landau gauge used in Ref.\ \onlinecite{Kita98-2}.
Hence there is an extra factor 
due to the gauge transformation
in every expression
of the basis functions,
such as ${\rm e}^{-{\rm i}xy/2 l_{\! B}^{2}}$
in Eq.\ (\ref{basis-k}) below.

The basis function
$\psi_{N{\bf k}\alpha}$ $(N\!=\! 0,1,2,\cdots; \,\,\alpha\!=\! 1,2)$ 
in Eq.\ (\ref{Dexp1}) is defined by
\begin{eqnarray}
\psi_{N{\bf k}\alpha}({\bf r})\!=&& \!\!\!\!\!\!\!\!\!
\sum_{n=-{\cal N}_{{\rm f}}/2+1}^{{\cal N}_{%
{\rm f}}/2}\!\!\!\!{\rm e}^{i [k_{y}(y+l_{\! B}^{2}k_{x}/2)
+na_{1x}(y+l_{\! B}^{2}k_{x}-na_{1y}/2)/l_{\! B}^{2}]}
\nonumber \\
&&\times{\rm e}^{-i xy/2l_{\! B}^{2}-(x\!- l_{\! B}^{2}k_{y}
- na_{1x})^{2}/2l_{\! B}^{2}+i(\alpha-1)n\pi}
\nonumber \\
&&\times \sqrt{\frac{a_{1x}/l_{\! B}}{2^{N}N!\sqrt{\pi }\,{\cal S} }}\,
H_{\! N}\!\!\left(\!
\frac{x\!-\! l_{\! B}^{2}k_{y}\! -\! na_{1x}}{l_{\! B}}\!\right) ,
\label{basis-k}
\end{eqnarray}
where ${\cal S}\!\equiv \!\pi l_{\! B}^{2}{\cal N}_{\rm f}^{2}$,
and $H_{N}(x)\!\equiv \! 
{\rm e}^{x^{2}} \left(-\frac{{\rm d}}{{\rm d}x}\right)^{\! N}
{\rm e}^{-x^{2}} $ is the Hermite polynomial.\cite{Abramowitz}
The basis function $\psi_{N{\bf q}}^{({\rm c})}$ for the center-of-mass coordinates
is obtained from Eq.\ (\ref{basis-k}) by putting 
${\bf k}\!\rightarrow\!{\bf q}$, $\alpha\!=\!1$,
and $l_{\! B}\!\rightarrow\! l_{{\rm c}}\!\equiv\! l_{\! B}/\sqrt{2}$
as
\begin{eqnarray}
\psi_{N{\bf q}}^{({\rm c})}({\bf r})\!=&& \!\!\!\!\!\!
\sum_{n=-{\cal N}_{{\rm f}}/2+1}^{{\cal N}_{%
{\rm f}}/2}\!\!\!\!{\rm e}^{i [q_{y}(y+l_{{\rm c}}^{2}q_{x}/2)
+na_{1x}(y+l_{{\rm c}}^{2}q_{x}-na_{1y}/2)/l_{{\rm c}}^{2}]}
\nonumber \\
&&\times{\rm e}^{-i xy/2l_{{\rm c}}^{2}-(x\!- l_{{\rm c}}^{2}q_{y}
- na_{1x})^{2}/2l_{{\rm c}}^{2}}
\nonumber \\
&&\times \sqrt{\frac{a_{1x}/l_{{\rm c}}}{2^{N}N!\sqrt{\pi }\,{\cal S} }}\, 
H_{\! N}\!\!\left(\!
\frac{x\!-\! l_{{\rm c}}^{2}q_{y}\! -\! na_{1x}}{l_{{\rm c}}}\!\right) .
\label{basis-c}
\end{eqnarray}
Finally, the basis function $\psi_{N m}$ for the relative coordinates, 
which is conveniently chosen as 
an eigenstate of the orbital angular momentum operator $\hat{l}_{z}$,
is given by
\begin{equation}
\psi_{N m}^{({\rm r})}({\bf r}) =\frac{(-1)^{N} }{ \sqrt{2\pi } \, l_{\rm r} } 
\sqrt{\!\frac{N!}{(N\! +\! m)!}}
\, \zeta^{m}
\, {\rm e}^{-|\zeta |^{2}/2} \, L^{(m)}_{N}(|\zeta|^{2}) \, ,
\end{equation}
where 
$\zeta\!\equiv\!(x\!+\!{\rm i}y)/\sqrt{2}l_{\rm r}$ with
$l_{{\rm r}}\equiv\sqrt{2}l_{\! B}$, and
$L^{(m)}_{N}(x)\!\equiv \! \frac{1}{N!}
{\rm e}^{x} x^{-m}\left(\frac{{\rm d}}{{\rm d}x}\right)^{\! N}
{\rm e}^{-x} x^{N+m}$ is the generalized Laguerre polynomial
satisfying $N+m\geq 0$.\cite{Abramowitz}

We next provide expressions of the overlap integrals in Eq.\ 
(\ref{D-Dt}).
The first one, which was obtaind by Rajagopal and Ryan,\cite{Rajagopal91} is given by
\begin{eqnarray}
&&\hspace{-3mm}\langle N_{{\rm c}}N_{{\rm r}}|N_{1}N_{2}\rangle
\nonumber \\
&&\hspace{-3mm}\equiv\delta_{N_{1}+N_{2},N_{\rm c}+N_{\rm r}}
\sqrt{\frac{N_{1}!N_{2}!N_{\rm c}!N_{\rm r}!}{2^{N_{1}+N_{2}}}}
\nonumber \\
&& \hspace{-1mm} \times\hspace{-2mm}
\sum_{n={\rm max}(0,N_{\rm c}-N_{2})}^{{\rm min}(N_{1},N_{\rm c})}
\frac{(-1)^{N_{2}-N_{\rm c}+n}}
{n!\,(N_{1}\!-\! n)!\,(N_{\rm c}\!-\! n)!\,(N_{2}\! - \! N_{\rm c}\! +\! n)!} \, ,
\nonumber \\
\label{overlap1}
\end{eqnarray}
which satisfies
\begin{subequations}
\label{overlap1prop}
\begin{equation}
\langle N_{{\rm c}}N_{{\rm r}}|N_{1}N_{2}\rangle
=(-1)^{N_{\rm r}}\langle N_{{\rm c}}N_{{\rm r}}|N_{2}N_{1}\rangle \, ,
\label{overlap1prop1}
\end{equation}
\begin{equation}
\sum_{N_{1}N_{2}}\langle N_{{\rm c}}'N_{{\rm r}}'|N_{1}N_{2}\rangle
\langle N_{1}N_{2}|N_{{\rm c}}N_{{\rm r}}\rangle =\delta_{N_{{\rm c}}'N_{{\rm c}}}
\, \delta_{N_{{\rm r}}'N_{{\rm r}}} \,  .
\label{overlap1prop2}
\end{equation}
\end{subequations}
The asymptotic expression of Eq.\ (\ref{overlap1}) for $N_{\rm c}\!
\ll\! N_{1},\! N_{2}$ is given by\cite{Kita98-2}
\begin{eqnarray}
&& \hspace{-3mm} \langle N_{{\rm c}}N_{{\rm r}}| N_{1}N_{2}\rangle 
\approx
 \delta_{N_{1}+N_{2},N_{\rm c}+N_{\rm r}}
\, (-1)^{N_{2}} \left[\frac{2}{\pi (N_{{\rm c}}\!+\! N_{{\rm r}})}\right]^{\! 1/4}
\nonumber \\
&&\hspace{22mm}\times
{\rm e}^{-{x^{2}}/{2}} \, \frac{H_{N_{\rm c}}(x)}{\sqrt{2^{N_{\rm c}}N_{\rm c}!}}
 \, ,
\label{B_NN'asym}
\end{eqnarray}
with $x\!\equiv\!(N_{1}\!-\!N_{2})/
\sqrt{2(N_{\rm c}\!+\! N_{\rm r})}$.
This expression forms the basis for
quasiclassical approximations.

The second overlap integral is given by
\begin{equation}
\langle {\bf q}| m\!+\!N_{\rm r}\rangle
= \sqrt{2\pi}\, l_{\rm r} \, 
(-1)^{ m+N_{\rm r}} \bigl[\psi_{ m+N_{\rm r}{\bf q}}^{({\rm r})}({\bf 0})\bigr]^{\! *} \, ,
\label{overlap2}
\end{equation}
where $\psi_{ m+N_{\rm r}{\bf q}}^{({\rm r})}$ 
is another basis function of the relative coordinates
defined by
\begin{eqnarray}
\psi_{N{\bf q}}^{({\rm r})}({\bf r})\!=&& \!\!\!\!\!\!\!\!\!\!
\sum_{n=-{\cal N}_{{\rm f}}/4+1}^{{\cal N}_{%
{\rm f}}/4}\!\!\!\!{\rm e}^{i [q_{y}(y+l_{{\rm r}}^{2}q_{x}/4)/2
+2na_{1x}(y+l_{{\rm r}}^{2}q_{x}/2-na_{1y})/l_{{\rm r}}^{2}]}
\nonumber \\
&&\times{\rm e}^{-i xy/2l_{{\rm r}}^{2}-(x\!- l_{{\rm r}}^{2}q_{y}/2
- 2na_{1x})^{2}/2l_{{\rm r}}^{2}}
\nonumber \\
&&\times \sqrt{\frac{2a_{1x}/l_{{\rm r}}}{2^{N}N!\sqrt{\pi }\,{\cal S} }}\, 
H_{\! N}\!\!\left(\!
\frac{x\!-\! l_{{\rm r}}^{2}q_{y}/2\! -\! 2na_{1x}}{l_{{\rm r}}}\!\right) .
\nonumber \\
\label{basis-r}
\end{eqnarray}

\section{\label{App:dHvA-analytic}Analytic Formula for dHvA oscillation damping}

In order to derive an
analytic expression for the dHvA oscillation damping 
in the vortex states, we start from
the thermodynamic potential of Eq.\ (\ref{Omega}).
The last term $-\frac{1}{2}\mbox{Tr}\underline{\Delta}\,\underline{\Phi}$ 
may be expressed solely with respect to  the pair potential, 
so that they can be neglected in the present model
to consider the oscillatory part.
Since we are interested in the extra damping in the vortex state,
we adopt as $E_{s}$ and ${\bf v}_{s}$ 
the expressions from the second-order perturbation with respect 
to $\underline{\Delta}$. They are obtained as
\begin{equation}
E_{N{\bf k}\alpha p_{z}\sigma}  = |\xi_{N p_{z}\sigma}|
\! +\! \eta_{N{\bf k}p_{z}}^{(1)}{\rm sign}(\xi_{Np_{z}\sigma}) \, ,
\label{E2nd}
\end{equation}
\begin{equation}
\int\!\! |{\bf v}_{N{\bf k}\alpha p_{z}\sigma}({\bf r})|^{2} 
d{\bf r} = \theta(-\xi_{Np_{z}\sigma})
\! +\! \eta_{N{\bf k}p_{z}}^{(2)} {\rm sign}(\xi_{Np_{z}\sigma}) \, ,
\label{vInt}
\end{equation}
where 
$\theta$ is the step function,
and $\eta_{N{\bf k}p_{z}}^{(n)}$ is defined by
using Eq.\ (\ref{D-Dt1}) as
\begin{eqnarray}
\eta_{N{\bf k}p_{z}}^{(n)}\equiv\sum_{N'}
\frac{|{\Delta}^{({\bf k}p_{z})}_{N N'}|^{2}}{
(\xi_{Np_z}\!+\!\xi_{N'\! -p_{z}})^{n}} .
\end{eqnarray}
The first terms on the right-hand side of Eqs.\ (\ref{E2nd}) and (\ref{vInt}) 
are just the normal-state results.
The second terms, on the other hand,
denote the finite quasiparticle dispersion 
in the magnetic Brillouin zone
and the smearing of the Fermi surface, respectively,
due to the scattering by the growing pair potential.
It is useful
to express $\eta_{N{\bf k}p_{z}}^{(n)}$ in terms of 
$\tilde{\Delta}^{\!(0)\!}(B)$ and the cyclotron energy $\hbar\omega_{B}$ 
of the extremal orbit as
\begin{eqnarray}
\eta_{N{\bf k}p_{z}}^{(n)} =
\frac{|\tilde{\Delta}^{\!(0)\!}(B)|^{2}}{(\hbar \omega_{B})^{n}} \,
\tilde{\eta}_{N{\bf k}p_{z}}^{(n)}
 \, .
\label{eta}
\end{eqnarray}
The quantity $\tilde{\eta}_{N{\bf k}p_{z}}^{(n)}$ thus defined 
is dimensionless, 
and we realize that the main $B$ dependence in Eq.\ (\ref{eta})
lies in the prefactor $|\tilde{\Delta}^{\!(0)\!}(B)|^{2}/(\hbar \omega_{B})^{n}$.
The explicit expression of $\tilde{\eta}_{N{\bf k}p_{z}}^{(n)}$ is obtained by
using Eqs.\ (\ref{D-Dt1}) and (\ref{pair3D}).
Considering the case $\theta_{\bf c}\!=\!0$, for simplicity, 
it is given by
\begin{eqnarray}
\tilde{\eta}_{N{\bf k}p_{z}}^{(n)} \! =  &&\hspace{-3mm}
\frac{{\cal N}_{\rm f}^{2}}{4}\!\! \sum_{N'mm'}\!\!
\frac{|\langle NN'| 0 \, N\!+\!N'\rangle|^{2}
\langle N\!+\!N'\!+\! m| 2{\bf k}\!-\!{\bf q} \rangle}{
[N\!+\! N'\!-\!2(N_{\rm F}\!+\!\delta)]^{n}} 
\nonumber \\
&& \hspace{-5mm}\times
\langle 2{\bf k}\!-\!{\bf q} |  N\!+\!N'\!+\! m' \rangle  \times 
\left\{
\begin{array}{l}
\vspace{1mm} \delta_{m0}\delta_{m'0}\\
\vspace{1mm} \delta_{m,\pm 2}\delta_{m',\pm 2}\sin^{4}\!\theta_{\bf p} \\
\delta_{m0}\delta_{m'0}\cos^{2}\!\theta_{\bf p}
\end{array}
\right. ,
\nonumber \\
\label{etaT}
\end{eqnarray}
where
the quantity $\delta\!= \!\delta(B,p_{z})$ ($|\delta|\!<\! 1/2$) specifies 
the location of $\varepsilon_{\rm F}$ between the two closest Landau levels,
and  the overlap integrals are defined by Eqs.\ (\ref{overlap1}) and (\ref{overlap2}).
The corresponding normalized density of states:
\begin{eqnarray}
D_{\hspace{-0.2mm}Np_{z}}^{(n)}\hspace{-0.6mm} (\tilde{\eta})
\equiv \frac{2}{{\cal N}_{\rm f}^{2}}\sum_{{\bf k}\alpha}
\delta(\,\tilde{\eta} -\tilde{\eta}_{N{\bf k}p_{z}}^{(n)})\, ,
\label{gDef}
\end{eqnarray}
will play a central role in the following.

Substituting Eqs.\ (\ref{E2nd}) and (\ref{vInt}) into Eq.\ (\ref{Omega}),
we find that the terms containing $\eta_{N{\bf k}p_{z}}^{(2)}$ may be
neglected due to the cancellation between the particle and hole 
contributions.
The remaining term can be transformed with the standard procedure.\cite{LK55}
We thereby obtain, for the first harmonic of
$\Omega/V$, the expression:
\begin{eqnarray}
\hspace{-4mm}
\frac{\Omega_{1}}{V} &&\hspace{-4mm} =  -\frac{k_{\rm B}T}{2\pi^{2} l_{B}^2} 
\sum_{\sigma}\int_{-1/2}^{\infty} \! \! dN \cos(2\pi N)\!
\int_{-\infty}^{\infty}\!\!dp_{z}\! \int_{-\infty}^{\infty} \!\! d\tilde{\eta} 
\nonumber \\
&&\hspace{-7mm}\times 
D_{\hspace{-0.2mm}Np_{z}}^{(1)}\hspace{-0.6mm} (\tilde{\eta}) 
\ln \bigl[1\!+\!{\rm e}^{-(\xi_{Np_{z}\sigma}
+\tilde{\eta}\,|\tilde{\Delta}^{\!(0)\!}(B)|^{2}/\hbar \omega_{B} )/k_{\rm B}T}
\hspace{0.2mm}\bigr] \, .
\label{Omega3}
\end{eqnarray}
The function $D_{\hspace{-0.2mm}Np_{z}}^{(1)}\hspace{-0.6mm}(\tilde{\eta})$ 
depends on $(N,p_{z})$,
but may be replaced by a representative one ${\overline D}_{\ell}^{(1)}\!(\tilde{\eta})$
to be placed outside the $N$ and $p_{z}$ integrals,\cite{comment2}
where the recovered index $\ell$ specifies the $s$-, $d$-, or $p$-wave case
of Eq.\ (\ref{etaT}).
It may also be acceptable to use a Lorenzian for it:
${\overline D}_{\ell}^{(1)}\!(\tilde{\eta})\!=\!
\tilde{\Gamma}_{\ell}/\pi(\tilde{\eta}^{2}\!+\!\tilde{\Gamma}_{\ell}^{2})$.\cite{comment3}
We thereby obtain an expression for the magnetization
which carries an extra damping factor:
\begin{eqnarray}
\hspace{-2mm}R_{\rm s}(B)  &&\hspace{-3mm} \equiv \int_{-\infty}^{\infty}\!\!
{\overline D}_{\ell}^{(1)}\!(\tilde{\eta})\exp\!\left[-2\pi i\hspace{0.2mm}
\tilde{\eta}\hspace{0.2mm}
|\tilde{\Delta}^{\!(0)\!}(B)|^{2}/(\hbar \omega_{B})^{2} \hspace{0.2mm}
 \right] d\tilde{\eta} 
\nonumber \\
&&\hspace{-3mm} = \exp\bigl[-2\pi\tilde{\Gamma}_{\ell}
|\tilde{\Delta}^{(0)\!}(B)|^{2}/(\hbar \omega_{B})^{2} \,\bigr] \, .
\label{R_D}
\end{eqnarray}
Thus, the superconductivity gives rise to an extra Dingle temperature of 
$k_{\rm B}T_{\Delta}\equiv \tilde{\Gamma}_{\ell}\,
|\tilde{\Delta}^{\!(0)\!}(B)|^{2}/\pi\hbar \omega_{B}$,
or equivalently,
the extra scattering rate of $\tau_{\rm s}^{-1}\!\equiv\!
2\pi k_{\rm B}T_{\Delta}/\hbar$.

Equation (\ref{R_D}) has an advantage that one
can trace the origin of the extra dHvA damping
definitely to the growing pair potential,
which brings finite quasiparticle dispersion 
in the magnetic Brillouin zone as Eq.\ (\ref{etaT}),
and the corresponding Landau-level broadening as Eq.\ (\ref{gDef}).
Moreover, Eq.\ (\ref{etaT}) reveals that this broadening near $H_{c2}$ is 
closely connected with the zero-field gap structure given by Eq.\ (\ref{gap3D}).

There seems to be no analytic way to estimate $\tilde{\Gamma}_{s}$, so
we fix it through the best fit to the numerical data of Fig.\ \ref{fig:dHvA-s}(b).
Using Eq.\ (\ref{Del(0)}) with $a^{2}\!=\!0.5\Delta_{0}^{2}$,
the procedure yields $
\tilde{\Gamma}_{s}\!=\!0.125$, as mentioned before.
It is also clear both from Figs.\ \ref{fig:dHvA-d} and \ref{fig:dHvA-p} 
and from Eq.\ (\ref{etaT}) that 
the average gap around the extremal orbit 
is relevant for the extra attenuation.
We hence put $a^{2}\tilde{\Gamma}_{\ell}\!=\!0.5
\langle|\Delta_{\bf p}|^{2}\rangle_{\rm eo}\tilde{\Gamma}_{s}$.
We thereby obtain Eq.\ (\ref{T_D}),
which yields excellent fits to the $d$- and $p$-wave 
numerical data without any adjustable parameters, as
seen in Figs.\ \ref{fig:dHvA-d} and \ref{fig:dHvA-p}.





\begin{thebibliography}{99}
\bibitem{comment1}This fact may be realized by observing that 
there is a clear zero-bias peak at the core (Fig.\ 8a) 
as well as the BCS gap structure (Fig.\ 8d) in their
density of states. The sharp peaks in Fig.\ 8 may be 
due to a very small value of $p_{\rm F}\xi_{0}$ $(\sim 1)$
in their model.
\bibitem{comment2}The $p_{z}$ integral in Eq.\ (\ref{Omega3})
passes through regions where $\delta\!\approx\!0$ in Eq.\ (\ref{etaT}), 
i.e., the second-order perturbation is not justified.
However, the main contribution to the integral certainly comes from the regions
where we can use Eq.\ (\ref{E2nd}). 
Notice that $\delta\!=\! -1/8$ at the maximal orbit 
when the magnetization takes a local maximum.
\bibitem{comment3}It is found numerically that the variance of Eq.\ (\ref{gDef}) 
depends little on the values of $N_{\rm F}$ and $N$ $(\sim\! N_{\rm F})$,
as expected.
It is $\sim\!0.2$ for the $s$-wave pairing with $\delta\!=\!-1/8$.
\end{thebibliography}
\end{document}